\documentclass[usenatbib]{mn2e}

\usepackage[dvips]{graphicx}
\usepackage{url}
\usepackage{caption}
\usepackage{footnote}
\usepackage{booktabs}
\usepackage{threeparttable}
\usepackage{epsfig}
\usepackage{cite}
\usepackage{hyperref}
\usepackage{breakurl}

\begin{document}

\title[Galaxy pre-processing in clusters at z$\sim$0.4]
{Galaxy pre-processing in substructures around z$\sim$0.4 galaxy clusters
\thanks {Based on data collected by the CLASH-VLT Large Programme 186.A-0.798; \citealt{rosati}), at the NASA HST, and at the NASJ SUBARU telescope
{\dag} : dolave@astro-udec.cl}
}
\author[D. Olave-Rojas et al.]
{
\parbox[t]{\textwidth}{
 D. Olave-Rojas$^{1}$\dag, P. Cerulo$^{1}$, R. Demarco$^{1}$, Y. L. Jaff\'{e}$^{2,3}$, A. Mercurio$^{4}$, P. Rosati$^{5}$, I. Balestra$^{6,7}$, M. Nonino$^{6}$}
\vspace*{6pt} \\
$^{1}$ Department of Astronomy, Universidad de Concepci\'{o}n, Casilla 160-C, Concepci\'{o}n, Chile\\
$^{2}$ Instituto de F\'{i}sica y Asatronom\'{i}a, Universidad de Valpara\'{i}so, Avda. Gran Breta\~{n}a 1111, Valpara\'{i}so, Chile \\
$^{3}$ European Southern Observatory, Alonso de Cordova 3107, Vitacura, Casilla 19001, Santiago de Chile, Chile\\
$^{4}$ INAF - Osservatorio Astronomico di Capodimonte, Via Moiariello 16 I-80131 Napoli, Italy\\
$^{5}$ Dipartimento di Fisica e Scienze della Terra, Universit$\grave{a}$ di Ferrara, Via Saragat 1, I-44122 Ferrara, Italy\\
$^{6}$ INAF - Osservatorio Astronomico di Trieste, via G. B. Tiepolo 11, I-34143, Trieste, Italy\\
$^{7}$ University Observatory Munich, Scheinerstrasse 1, D-81679 Munich, Germany\\
}
\date{\today}

\pagerange{\pageref{firstpage}--\pageref{lastpage}}

\maketitle

\label{firstpage}

\begin{abstract}

We present a detailed analysis of galaxy colours in two galaxy clusters at \mbox{z $\sim$ 0.4}, \mbox{MACS J0416.1-2403} and \mbox{MACS J1206.2-0847}, drawn from the CLASH-VLT survey, to investigate the role of pre-processing in the quenching of star formation. We estimate the fractions of red and blue galaxies within the main cluster and the detected substructures and study the trends of the colour fractions as a function of the projected distance from the cluster and substructure centres. Our results show that the colours of cluster and substructure members have consistent spatial distributions. In particular, the colour fractions of galaxies inside substructures follow the same spatial trends observed in the main clusters. Additionally, we find that at large cluster-centric distances \mbox{($r \geq r_{200}$)} the fraction of blue galaxies in both the main clusters and in the substructures is always lower than the average fraction of UVJ-selected star-forming galaxies in the field as measured in the COSMOS/UltraVista data set. We finally estimate environmental quenching efficiencies in the clusters and in the substructures and find that at large distances from the cluster centres, the quenching efficiency of substructures becomes comparable to the quenching efficiency of clusters. Our results suggest that pre-processing plays a significant role in the formation and evolution of passive galaxies in clusters at low redshifts.

\end{abstract}


\section{Introduction}\label{intro}

It is well established that the evolution of galaxies is driven by a combination of internal and external physical mechanisms,  which are linked to their stellar mass and to the environment in which they reside, respectively. In particular, the question: how do the properties of galaxies change as a function of environment? has motivated many works in the last decades (e.g. \citealt{dressler80}, \citealt{couch98}; \citealt{kauffmann04}, \citealt{postman05}, \citealt{baldry06}, \citealt{bravo09}, \citealt{peng10}, \citealt{jaffe16}, \citealt{nantais17}, \citealt{cerulo17}). 

Galaxy clusters are the most massive gravitationally bound cosmic structures and an ideal laboratory for the study of the environmental drivers of galaxy evolution (e.g. \citealt{poggianti06}, \citealt{delucia07}, \citealt{demarco10}, \citealt{lemaux}, \citealt{cerulo14}). The broad range of densities available in these systems, ranging from the dense core to the sparse outskirts, allows one to study the different physical mechanisms responsible for galaxy transformations (e.g. \citealt{boselli16}).

According to the $\Lambda$ cold dark matter ($\Lambda$CDM) hierarchical paradigm, galaxy clusters are assembled through the continuous merging of smaller, group-like structures (\citealt{press74}, \citealt{fakhouri10}, \citealt{chiang13}). Theoretical models show that massive clusters in the Local Universe, with dark matter halo masses of the order of \mbox{$10^{14.5} - 10^{15} M_{\odot}$}\footnote{We converted all the quantities reported in the literature to the cosmological parameters adopted in this study.\label{note1}}, accreted \mbox{$\sim$ 40\%} of their galaxies from infalling groups with masses of the order of \mbox{$10^{12} - 10^{13} M_{\odot}$\textsuperscript{\ref{note1}}} (\citealt{mcgee}). For this reason, the study of the properties of galaxies at large cluster-centric radii (\mbox{$2 \times r_{200}$ $< r < 3 \times r_{200}$}\footnote{$r_{200}$ is the typical radius of a sphere with a mean density equal to 200 times the critical density}; e.g. \citealt{li09}, \citealt{valentinuzzi}, \citealt{lemaux}, \citealt{dressler13}, \citealt{hou}, \citealt{just}, \citealt{haines15}), where these systems are still in the process of assembling, is of primary importance to understand the connection between the evolution of galaxies and the formation of their hosting large-scale structures (\citealt{fitchett88a}, \citealt{eke96}, \citealt{kravtsov}). 

In particular, some authors (see e.g. \citealt{vijayaraghavan}, \citealt{hou}; \citealt{just}; \citealt{haines15}) find that the infall regions of galaxy clusters, at cluster-centric distances \mbox{$2 - 3 \times r_{200}$}, host large fractions of quiescent galaxies, with little or no ongoing star formation. This result cannot be reproduced by theoretical models in which star-formation, in infalling field galaxies, is quenched only when the galaxies pass within $r_{200}$ of the galaxy cluster \citep{haines15}. In order to reproduce the observational results, it is necessary that the star formation in those galaxies be quenched in groups prior to the infall into the cluster. This scenario is known as \textit{pre-processing} (\citealt{zabludoff}, \citealt{fujita}, \citealt{wetzel13}).

Theoretical studies (e.g \citealt{fujita}, \citealt{mcgee}, \citealt{vijayaraghavan}, \citealt{wetzel13}, \citealt{bahe13}, \citealt{bahe15}) suggest that pre-processing is responsible for the elevated fraction of quiescent galaxies observed in the outer regions of galaxy clusters (e.g. \citealt{hou}, \citealt{haines15}), in comparison with the field. In particular, \citet{vijayaraghavan} performed cosmological N-body simulations and hydrodynamic simulations to study the merger between a galaxy group and a galaxy cluster since \mbox{z = 0.5}. The results of this analysis predict that pre-processing plays a fundamental role in quenching star formation because the gas of infalling group members is removed inside galaxy groups before these groups fall into a galaxy cluster. The gas is removed mainly through ram-pressure stripping \citep{gunn72} and galaxy-galaxy interactions (e.g. \citealt{toomre}, \citealt{barnes96}).  Moreover, these authors, following the cluster-group merger from \mbox{$z = 0.5$} to \mbox{$z = 0$}, show that the infalling group does not get immediately viralised in the cluster environment: at \mbox{$z = 0.2$} it can be seen that there are substructures within the main body of the cluster containing traces of the kinematics of the group. Similarly, \citeauthor{bahe13} (\citeyear{bahe13}, \citeyear{bahe15}) using a high-resolution cosmological hydrodynamic simulations found that \mbox{$\sim$ 50\%} of galaxies in subhaloes near to a massive galaxy cluster have been affected by pre-processing. 

If the predictions of \citet{vijayaraghavan} are correct then it should be possible to observe galaxy properties in cluster substructures that are not yet virialised within the main cluster halo. \citet{hou} studied the impact of pre-processing through the observed quenched fraction in a sample of groups and cluster galaxies from the Sloan Digital Sky Survey Data Release 7 (SDSS-DR7; \citealt{abazajian09}) in the redshift range of \mbox{$0.01 < z < 0.045$}. They applied the Dressler-Shectman statistic (also known as the Dressler-Shectman test or DS-test, \citealt{dressler88}) to identify substructures which are kinematically distinct from the main galaxy cluster \citep{dressler13}. In particular, \citet{hou} found that at \mbox{$2 \times r_{200} < r < 3 \times r_{200}$}, the fraction of quiescent galaxies is higher in the substructure population than in the field population. These authors suggest that this enhancement is a result of the pre-processing of galaxies within substructures. In the same context, \citet{cybulski} analysed the effects of pre-processing in a sample of 3505 galaxies in the Coma Supercluster. They studied the star formation (SF) activity of galaxies as a function of the type of environment (e.g. cluster, group, filament and void) to quantify the degree of impact of the environment on the SF activity in galaxies. They found that the pre-processing plays a fundamental role at low redshift, and that the evolution driven by the environment affects \mbox{$\sim$ 50\%} of the galaxies in groups.

\citeauthor{jaffe11b} (\citeyear{jaffe11b}, \citeyear{jaffe16}) found that galaxies within substructures are more likely to be deficient in atomic hydrogen (H{\sc i}) and passive. In addition, \citet{haines15} found that the fraction of star-forming cluster galaxies rises steadily from the centre to the outskirts of galaxy clusters, but even at $3 \times r_{200}$ the values remain \mbox{20 - 30\%} below field values. To explain these results it is necessary that the star-formation and the fraction of H{\sc i} in galaxies decline for the first time outside the central cluster regions, probably during the infall of galaxy groups.

In this paper we present a detailed analysis of the properties of galaxies in two clusters at \mbox{$z \sim 0.4$}, namely MACS J0416.1-2403 and MACS J1206.2-0847 (hereafter MAC0416 and MACS1206, respectively), drawn from Dark Matter Mass Distributions of Hubble Treasury Clusters and the Foundations of $\Lambda$CDM Structure Formation Models survey (CLASH-VLT, \citealt{rosati}). These two systems represent two extreme cases of galaxy clusters, MACS0416 being an ongoing merger of two clusters (see e.g. \citealt{balestra16}) and the second system, MACS1206, being a relaxed cluster \citep{biviano}. This paper is the first in a series addressing the study of the properties of galaxies in the substructures of clusters at intermediate redshifts (\mbox{$0.2 < z < 0.6$}) in the CLASH-VLT survey. We present here the method that will be used in the analysis of CLASH-VLT and apply it to two extreme examples of galaxy clusters.

The paper is organised as follows. In Section \ref{data_section} we describe the data sets. In Section \ref{analysis_section} we present the measurements and data analysis, while in Section \ref{result_section} we present the results which we discuss in Section \ref{discussion}. Section \ref{conslusion} summarises our main conclusions. Throughout the paper we adopt a $\Lambda$CDM cosmology with \mbox{$\Omega_{\Lambda} = 0.7$}, \mbox{$\Omega_{m} = 0.3$}, and \mbox{$h = H_0/100~km~s^{-1}~Mpc^{-1} = 0.7$} \citep{spergel}. All magnitudes in this paper are in the AB-system \citep{oke} unless otherwise stated. In this paper we consider $r_{200}$ as the physical radius, measured in Mpc, inside which the density is 200 times the critical density of the Universe at the redshift of each cluster.

\section{DATA}\label{data_section}

MACS0416 and MACS1206 were drawn from the CLASH-VLT survey. CLASH-VLT is a spectroscopic follow-up of the Cluster Lensing and Supernova survey with Hubble (CLASH, \citealt{postman12}), which targeted 25 massive clusters (\mbox{$M_{halo} > 10^{14.5} M_\odot$}) at redshifts \mbox{$0.2 < z < 0.9$} with the Hubble Space Telescope and other space- and ground-based facilities to study cosmology and galaxy evolution. CLASH-VLT targeted the 13 CLASH clusters at redshifts \mbox{$0.2 < z < 0.6$} that are observable from the southern hemisphere. In this paper we use public spectrophotometric data compiled by the CLASH colaboration\footnote{\url{https://archive.stsci.edu/prepds/clash/}}.

\subsection{Photometric Catalogues}\label{photometric_catalogues}
MACS0416 and MACS1206 were observed with the SuprimeCam \citep{miyazaki} at the prime focus of the 8.3m Subaru Telescope, in the  $B$, $R_{c}$ and {\it z'}, and $B$, $V$, $R_{c}$, $I_{c}$ and {\it z'} bands, respectively (\citealt{umetsu12}, \citeyear{umetsu14}). Photometric data from Subaru have an angular coverage of \mbox{$34' \times 27'$} around the centre of each galaxy cluster. The above observations are available in the Subaru archive, Subaru-Mitaka Okayama-Kiso Archive System (SMOKA\footnote{\url{http://smoka.nao.ac.jp}}; \citealt{baba}). In addition, MACS0416 was observed with the Wide-Field Imager (WFI; \citealt{baade}) on the ESO/MPG 2.2m telescope at the La Silla Observatory in Chile, in the  B, V, R and I photometric bands, covering an area of \mbox{$34' \times 33'$} around the centre of the cluster \citep{gruen14}. Subaru images were reduced by the CLASH team following the techniques described in \citet{nonino}. WFI images were reduced by \citet{gruen14} using \texttt{Astro-WISE} \citep{valentijn}. A full description of the reduction of Subaru and WFI images is given in \citet{umetsu12} and \citeauthor{gruen13} (\citeyear{gruen13}, \citeyear{gruen14}), respectively.

In this paper we use the photometric catalogues generated by the CLASH-VLT team. For MACS0416 we complement those catalogues with the public catalogue published by \citet{gruen14} and available on the author's web page\footnote{\url{http://www.usm.uni-muenchen.de/~dgruen/download.html}\label{note5}}.

Aperture magnitudes in Subaru and WFI catalogues were obtained using the software \texttt{SExtractor} \citep{bertin96} in conjunction with \texttt{PSFEx} \citep{bertin11}. Aperture magnitudes were measured within fixed circular apertures. For MACS0416 aperture diameters of 3$''$ and 2$''$ are used for Subaru and WFI, respectively, while for MACS1206 is used a diameter of 5$''$. Aperture magnitudes were corrected for zero-point shifts and Galactic extinction by comparing the galaxy observed colours to the stellar library of \citet{pickles} and using the Galactic extinctions of \citet{shlafly} and \citet{schlegel} for Subaru and WFI, respectively. We select galaxies with magnitudes \mbox{R$_{c} < 26.0$}, which corresponds to the $3\sigma$ limit for an aperture 2$''$ in diameter \citep{umetsu14}. A description of the photometric catalogues from Subaru and WFI can be found in \citeauthor{mercurio14} (\citeyear{mercurio14}, \citeyear{mercurio16}) and \citeauthor{gruen13} (\citeyear{gruen13}; \citeyear{gruen14}), respectively.

\subsection{Spectroscopic Catalogues}\label{spectroscopic_catalogues}
We use the public spectroscopic redshift catalogues\footnote{\url{https://sites.google.com/site/vltclashpublic/} \label{note6}} (see \citealt{biviano}; \citealt{balestra16}; \citealt{caminha}) from CLASH-VLT \citep{rosati}. 

The VIMOS spectroscopic observations were made in four separate pointings centred on the core of each cluster. MACS0416 and MACS1206 were observed using a total of 21 (15 Low-Resolution Blue and 6 Medium Resolution) masks and 12 (8 Low-Resolution Blue and 4 Medium Resolution) masks, respectively. Low-Resolution Blue and Medium-Resolution masks have a spectral coverage of \mbox{3760 - 6700{\AA}} and \mbox{4800 - 10000{\AA}} with a resolving power of \mbox{R = 180} and \mbox{R = 850}, respectively. A full description about how the spectroscopic observations of MACS0416 and MACS1206 were made can be found in \citet{balestra16} and \citet{biviano}, respectively.

The VIMOS spectroscopic observations were reduced, by the CLASH-VLT team, using the VIMOS Interactive Pipeline Graphical Interface (VIPGI, \citealt{scodeggio}). Redshifts were determined using the software \texttt{Easy Redshift} (EZ, \citealt{garilli}). EZ determines the redshift by making a cross-correlation between the observed spectrum and template spectra. In cases in which the redshift solution was dubious, the redshift was determined by visual inspection \citep{balestra16}. Uncertainties on the redshifts are in the range of \mbox{$75 - 150~km~s^{-1}$} (\citealt{annunziatella14}; \citealt{balestra16}).

Spectroscopic catalogues, of MACS0416 and MACS1206, have a magnitude limit of \mbox{R$_{c} = 24.0$ mag}. A detailed description of the CLASH-VLT redshift catalogues can be found in \citet{biviano}, \citet{mercurio14}, \citet{balestra16}, and \citet{caminha}.

The CLASH-VLT data release provides a spectroscopic sample of 4594 and 2736 galaxies for MACS0416 and MACS1206, respectively (\citealt{biviano}, \citealt{balestra16}, \citealt{caminha}), over an area of \mbox{$26' \times 23'$} around the centre of each galaxy cluster. 

\section{MEASUREMENT AND DATA ANALYSIS}\label{analysis_section}
\subsection{Spectroscopic Completeness}\label{completeness_section}
Due to the strategy used on the selection of spectroscopic targets in the CLASH-VLT survey (see \citealt{biviano}, \citealt{annunziatella14}, \citealt{balestra16}, \citealt{caminha} for details on the strategy of spectroscopic observations) the spectroscopic completeness of the CLASH-VLT sample decreases with the distance from the cluster centre, reducing the statistics in the outskirts of galaxy clusters This may affect our ability to detect substructures \citep{balestra16}.

We estimated the sample spectroscopic completeness following the {\sc appendix} A in \citet{poggianti06}. More precisely, we measured the ratio between the number of galaxies in the spectroscopic and photometric catalogues ($N_{spec}$ and $N_{phot}$, respectively) in different bins of apparent magnitude or projected distance from the cluster centre to estimate the spectroscopic completeness, \mbox{$C = N_{spec}/N_{phot}$}, which is used to correct our measurements. We stress here that $N_{spec}$ corresponds to the number of galaxies with spectroscopic redshift (see \S \ref{spectroscopic_catalogues}) and N$_{phot}$ corresponds to the number of all galaxies in the photometric catalogue.

\begin{figure} 
       \includegraphics[width=\columnwidth]{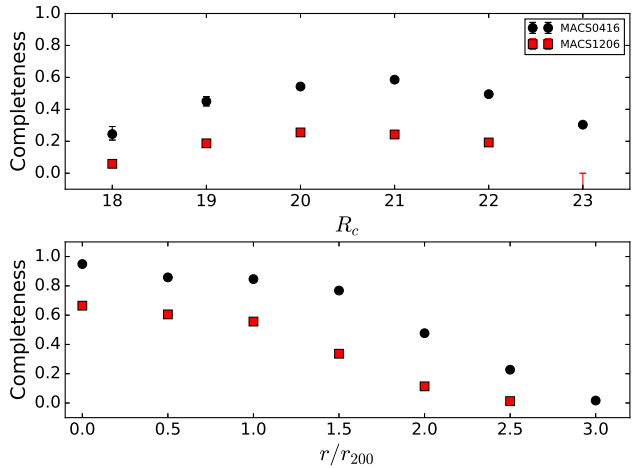}
	\caption{The top panel shows the completeness as a function of apparent magnitude in the R$_{c}$-band. The bottom panel shows the completeness as a function of the projected distance from the centre of the clusters normalised by $r_{200}$. Black points and red squares represent the spectroscopic completeness for MACS0416 and MACS1206, respectively. We note a lack of completeness in the brightest magnitude bins. This effect could be a consequence of the selection criteria for spectroscopic targets, which were selected mainly in a colour-colour space.}
	\label{fig1}
\end{figure}

In Figure \ref{fig1}, we show the spectroscopic completeness, for MACS0416 and MACS1206, as a function of galaxy apparent magnitude in the $R_{c}$-band and projected distance from the cluster centre normalised by $r_{200}$ ($r_{200}$ is 1.82 Mpc and 2.33 Mpc for MACS0416 and MACS1206, respectively; see \citealt{umetsu14}, \citealt{merten}). The central coordinates of each galaxy cluster were taken from \citet{balestra16} and \citet{annunziatella14} for MACS0416 and MACS1206, respectively.

\subsection{Photometric Redshifts}\label{zphot_section}
With the spectroscopic incompleteness increasing towards the cluster outskirts, we complemented our spectroscopic selection of cluster members and substructures with the photometric redshifts available for the two clusters. For MACS0416 we used the catalogue of \citet{gruen14}, released as part of the Hubble Frontier Fields (HFF\footnote{\url{http://www.stsci.edu/hst/campaigns/frontier-fields/}}, \citealt{koekemoer}, \citealt{lotz17}), while for MACS1206 we used an internal catalogue produced by the CLASH-VLT collaboration.

Photometric redshifts (hereafter $z_{phot}$) for MACS1206 were estimated using aperture magnitudes (see \S \ref{photometric_catalogues}) through a Neural Network method using the Multi Layer Perceptron with Quasi Newton Algorithm (MLPQNA; \citealt{brescia13}), which is an empirical method that achieves a high accuracy and reduces the number of catastrophic objects \citep{brescia13}. A full description of the derivation of $z_{phot}$ for MACS1206 is given in \citet{biviano} and \citet{mercurio14}. Photometric redshifts for MACS0416 were estimated using the photometric template-fitting developed by \citet{bender}. See \citet{brimioulle} and \citeauthor{gruen13} (\citeyear{gruen13}; \citeyear{gruen14}), for further details on $z_{phot}$ estimates for MACS0416.

\section{RESULTS}\label{result_section}
\subsection{Cluster membership, velocity dispersion and stellar mass}\label{spec_members}
It is crucial to determine the cluster membership to find and characterise substructures. In practice, cluster members are defined as those galaxies with a peculiar velocity (see equation \ref{peculiar_velocity}) lower than the escape velocity (v$_{esc}$, see equation \ref{escape_velocity}, \citealt{diaferio99b}), of a galaxy cluster.  

The peculiar velocity of a galaxy with redshift $z$ in the rest frame of a galaxy cluster with redshift $z_{cl}$ is given by:

\begin{equation}\label{peculiar_velocity}
v = c\frac{z-z_{cl}}{1+z_{cl}}
\end{equation}

This equation is valid, to first order, for \mbox{v $<<$ c} \citep{harrison}. The escape velocity, estimated for a cluster, using $M_{200}$ and $r_{200}$, is computed as \citep{diaferio99b}:

\begin{equation}\label{escape_velocity}
v_{esc} \simeq 927 \left(\frac{M_{200}}{10^{14}h^{-1}M_{\odot}}\right)^{1/2} \left(\frac{r_{200}}{h^{-1}Mpc}\right)^{-1/2} km s^{-1} 
\end{equation}

The estimates of $M_{200}$ using strong and weak gravitational lensing span the ranges \mbox{1.299 - 1.240 $\times 10^{15} M_{\odot}$} (see \citealt{umetsu14}, \citealt{umetsu16}) and \mbox{1.186 - 1.590 $\times 10^{15} M_{\odot}$} (see \citealt{biviano}, \citealt{umetsu14}, \citealt{merten}, \citealt{barreira15}) for MACS0416 and MACS1206 , respectively. The estimates of $M_{200}$ have a typical uncertainty \mbox{$\leq$ 5\%} depending of the model fitted (typically \citealt{einasto65} and \citealt{navarro95}) to the mass profile derived from gravitational lensing. Hereafter the values of $M_{200}$ assumed are \mbox{1.27 $\times 10^{15} M_{\odot}$} and \mbox{1.39 $\times 10^{15} M_{\odot}$} for MACS0416 and MACS1206, respectively. These values correspond to the mean value of $M_{200}$ for each cluster. Instead the mean value of $r_{200}$ to derive v$_{esc}$ was estimated using the equation (7) presented by \citet{finn05}.

Spectroscopic members were determined similarly to \citet{cerulo16}, first we removed galaxies from the redshift catalogue with a peculiar velocity higher than the v$_{esc}$ of the galaxy cluster. Then we estimated the velocity dispersion for the clean sample of cluster members, following the biweight estimator described in \citet{beers}, and finally we removed the field interlopers through a 3$\sigma$ clipping algorithm (see \citealt{yahil77}). Following the above criterion, we found \mbox{$\sim$ 890} and \mbox{$\sim$ 640} spectroscopic members for MACS0416 and MACS1206, respectively. We used the biweight estimator to estimate the mean cluster redshift and velocity dispersion of the whole system. We estimated mean redshifts of \mbox{$z \sim 0.397$} and \mbox{$z \sim 0.440$}, and velocity dispersions ($\sigma_{cl}$) of \mbox{1044 $\pm$ 23 km s$^{-1}$} and \mbox{1011 $\pm$ 25 km s$^{-1}$} for MACS0416 and MACS1206, respectively. These values are consistent within 1$\sigma$ and 1.4$\sigma$, respectively, with results reported in the literature (\citealt{girardi15}, \citealt{balestra16}). The errors in $\sigma_{cl}$ were estimated using a bootstrap technique. The central positions in Right Ascension (R.A.) and Declination (Dec.) and the kinematic properties of MACS0416 and MACS1206 are summarised in Table \ref{summarised_properties}

\begin{table*}
\centering
\begin{threeparttable}
	\caption{Main properties of the clusters MACS0416 and MACS1206}
  	\label{summarised_properties}
	{\small
 		\begin{tabular}{lccccc}
		\hline
		\hline
		Cluster    & R.A.\tnote{\emph{a}} & Dec.\tnote{\emph{a}} & v$_{esc}$     &  $z$    &  $\sigma_{cl}$   \\
		           &               \multicolumn{2}{c}{(J2000.)}  & Km s$^{-1}$   &         &  Km s$^{-1}$     \\ 
		\hline
		\hline
		MACS0416  &  04:16:09.14          &  -24:04:03.1         & 2375 $\pm$ 10 &  0.397  & 1044 $\pm$ 23    \\
		MACS1206  &  12:06:12.15          &  -08:48:03.4         & 2485 $\pm$ 68 &  0.440  & 1011 $\pm$ 25    \\
		\hline
		\hline
		\end{tabular}  
\begin{tablenotes}
\item[\emph{a}]{The centre positions were taken from \citet{balestra16} and \citet{annunziatella14} for MACS0416 and MACS1206, respectively.}
\end{tablenotes}
}
\end{threeparttable}
\end{table*}

\subsection{Detection of Substructures}\label{ds_substructures}

Once we have selected the spectroscopic members for each galaxy cluster, we verified whether the clusters contain substructures or not. For this, we used the Dressler-Shectman's statistic \citep{dressler88}, which allows us to make a test (also known as the Dressler-Shectman test or DS-test) to verify the existence of regions kinematically distinct from the main galaxy cluster \citep{dressler13}. 

In short, the DS-test compares the local velocity and velocity dispersion for each galaxy ($\bar{v}^{i}_{local}$ and $\sigma^{i}_{local}$) with the cluster global values ($\bar{v}_{cl}$ and $\sigma_{cl}$; \citealt{jaffe13}). Local parameters are estimated in a subset of galaxies containing the galaxy $i$ and its nearest neighbours ($N_{nn}$). We used \mbox{N$_{nn} = 10$} in equation \ref{delta_i} to apply the DS-Test in our clusters. The $\delta_{i}$ statistic used in the DS-Test is expressed as: 

\begin{equation}\label{delta_i}
\delta_{i}^{2} = \left(\frac{N_{nn} + 1}{\sigma_{cl}^{2}}\right)[(\bar{v}^{i}_{local} - \bar{v}_{cl})^{2} + (\sigma^{i}_{local} - \sigma_{cl})^{2}]~.
\end{equation}

This value quantifies the galaxy's kinematic deviation with respect to the mean cluster values of velocity and velocity dispersion (e.g. \citealt{dressler13}, \citealt{jaffe13}, \citealt{hou}). The larger the $\delta_{i}$ value, the greater the probability that the galaxy belongs to a substructure. 

\citet{dressler88} also define the cumulative $\delta$ as \mbox{$\Delta = \sum_i \delta_i$}. A value \mbox{$\Delta$/N$_{mem}$ $>$ 1}, where $N_{mem}$ is the number of cluster members, would be an indication that the cluster hosts substructures. However, it should be stressed that a high $\Delta$ value may be the result of random spatial fluctuations in the redshift distribution of cluster members. In order to assess this effect, we generated 1000 simulated spectroscopic samples by shuffling the velocities and positions of each galaxy in the clusters. For each of these bootstrap samples we estimated $\Delta$ and defined the P-value \mbox{P $= \sum(\Delta_{shuffle} > \Delta_{obs}/N_{shuffle})$} where $\Delta_{shuffle}$ is the value of $\Delta$ obtained for each simulated sample and $N_{shuffle}$ is the number of bootstrap iterations. Values of  \mbox{P $<$ 0.01} provide a robust constraint on the presence of substructures in the clusters. We find that MACS0416 and MACS1206 both have \mbox{$\Delta$/N$_{mem} >$ 1} and \mbox{P $<$ 0.001} (see Table \ref{Delta_statistics}), giving statistical support to the existence of real substructures.

\begin{table}
\centering
\begin{threeparttable}
	\caption{Results from the DS-Test on MACS0416 and MACS1206}
  	\label{Delta_statistics}
	{\small
 		\begin{tabular}{lccc}
		\hline
		\hline
 		Cluster     & N$_{mem}$ & $\Delta_{obs}/N_{mem}$ & P  \\
		\hline
		\hline
		MACS0416  &  890  & 1.64 & $<$0.001   \\
		MACS1206  &  641  & 1.22 & $<$0.001   \\
		\hline
		\hline
		\end{tabular}  
}
\end{threeparttable}
\end{table}

After applying the DS-test we proceeded to identify and characterise the substructures. For this purpose we need to select only galaxies with a \mbox{$\delta_{i}$ $>$ $\delta_{lim}$}, which corresponds to galaxies with a higher probability to be inside a substructure. To determine $\delta_{lim}$ we estimated the width $\sigma_{\delta}$ of the $\delta_{i}$ distribution in each cluster. Following \citet{girardi97}, $\delta_{lim}$ was defined as \mbox{$\delta_{lim}$ $=$ $3\sigma_{\delta}$} and galaxies were considered members of substructures if \mbox{$\delta_{i}$ $\geq$ $\delta_{lim}$}.

However, the selection \mbox{$\delta_{i}$ $\geq$ $\delta_{lim}$} may include galaxies which do not necessarily belong to a substructure due to their peculiar velocities and positions within the cluster (see e.g. discussions in \citealt{jaffe13}). Thus we added to the selection in $\delta_i$ two consecutive selections in peculiar velocity and projected position. The two selections were performed by using \texttt{python} scripts that combined new and pre-existing modules. The selection in velocity was performed, following e.g. \citet{girardi05} and \citet{demarco07} by using the Gaussian Mixture Model (GMM; \citealt{muratov}) algorithm. This algorithm assumes that a sample is described by the sum of two or more Gaussian functions. GMM estimates the probability that an object belongs to each identified Gaussian component through an iterative algorithm (expectation-maximisation, EM, \citealt{dempster}, \citealt{press}). The objects are assigned to the groups for which the likelihood of membership is higher.

For the selection of substructures in projected space, we tested two clustering algorithms available in \texttt{python}, namely \mbox{K-Means} \citep{lloyd} and the Density-Based Spatial Clustering of Applications with Noise (DBSCAN, \citealt{ester}), finally preferring the second one. To identify groups using DBSCAN we must define a minimum number of neighbouring objects separated by a specific distance. When using this algorithm not all objects in the sample are assigned to a group \citep{ester} and we can remove the galaxies that are not spatially grouped with others.

\begin{figure*} 
\centering
        \includegraphics[width=0.5\textwidth]{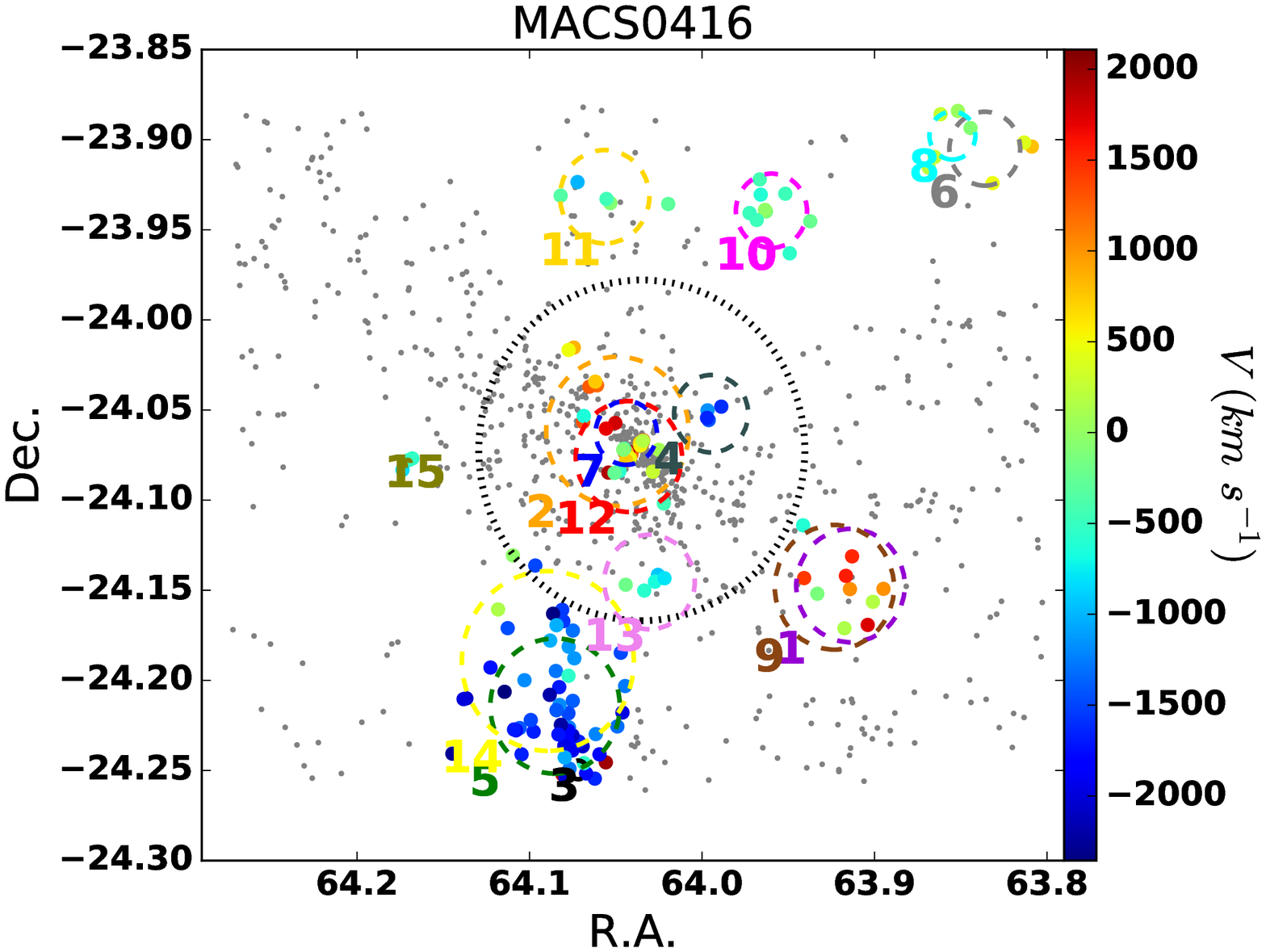}
        \hspace{-0.2cm}
        \includegraphics[width=0.5\textwidth]{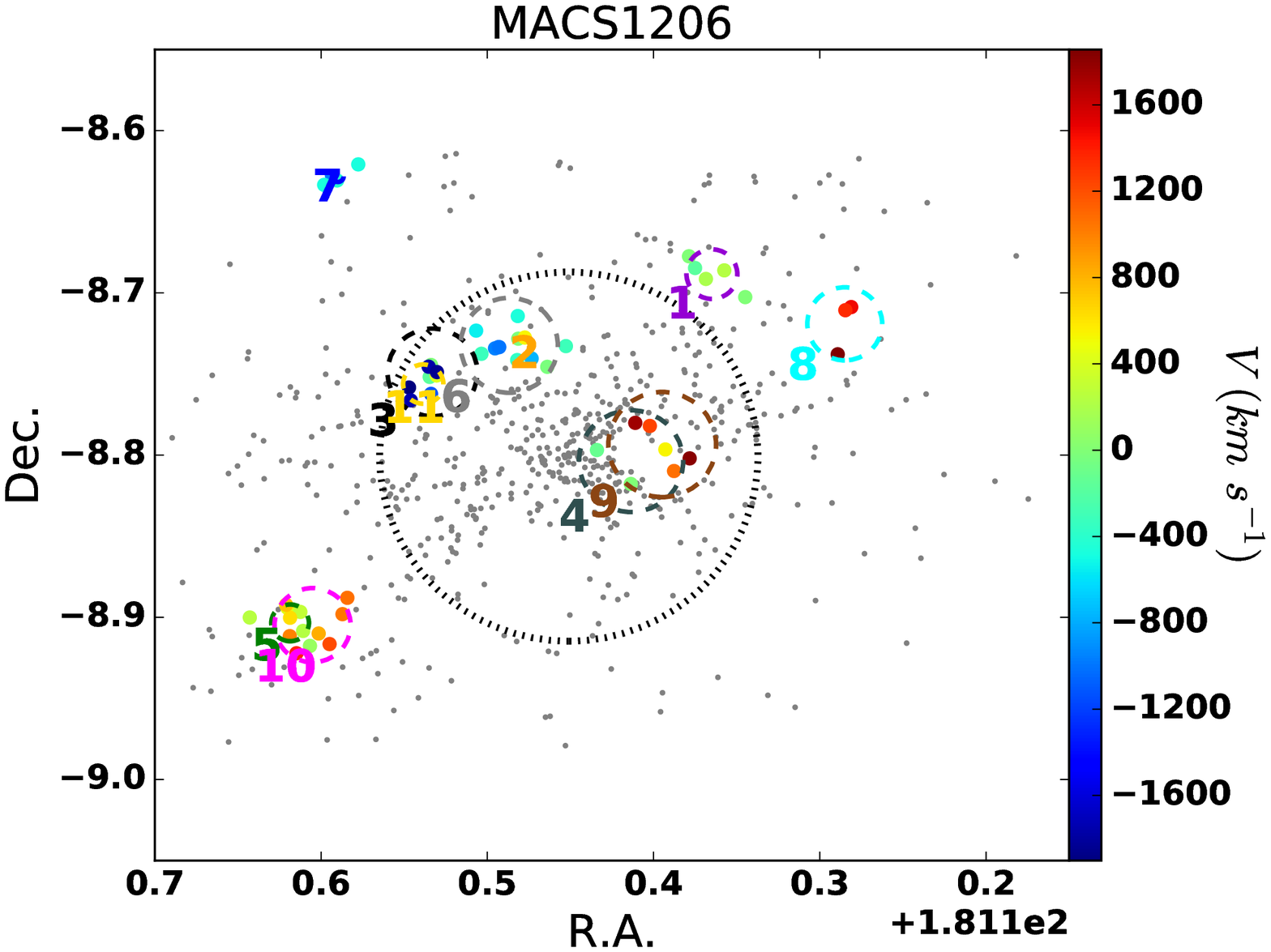}
	\caption{\small{Spatial and radial velocity distribution of cluster members for MACS0416 (left) and MACS1206 (right). The black dashed circle indicates the $r_{200}$ radius of the cluster. Grey points represent all spectroscopic members. Dashed circles of different colours indicate the $r_{200}$ of each identified substructure. Galaxies inside substructures are plotted with points whose colours indicate the peculiar velocity of the galaxies in the rest frame of the cluster.}}
	\label{substructures}
\end{figure*}

In practice, we defined a substructure as a collection of at least three neighbouring galaxies with a spatial separation of \mbox{$\sim$ 140 kpc}, which is a typical maximum spacing between galaxies within compact groups of galaxies \citep{sohn15}. Figure \ref{substructures} shows the substructures identified for each cluster. Table \ref{substructure_features} lists the central position, central redshift, number of members, velocity dispersion ($\sigma_{sb}$) and $r_{200}$ for each identified substructure. The central position of a substructure was defined as the centroid of the spatial distribution of galaxies in the substructure. $r_{200}$ was estimated from $\sigma_{sb}$ using Equation 8 of \citet{finn05} under the assumption that each substructure is virialised. Uncertainties on the numbers of substructure members were estimated as the 90\% Poissonian confidence intervals adopting the approximations of \citet{ebeling}. We note that MACS0416\_3, MACS0416\_15, MACS1206\_2 and MACS1206\_7 have a narrow velocity range and we cannot estimate $\sigma_{sb}$ and $r_{200}$ in these substructures. We found 15 substructures in MACS0416 and 11 in MACS1206 (see Figure \ref{substructures} and Table \ref{substructure_features}). 

We note that the substructure MACS0416\_5 corresponds to the \textit{``Sext''} substructure previously identified by \citet{balestra16} in MACS0416. For this substructure we estimate a number of members of \mbox{47 $\pm$ 10} and a velocity dispersion of \mbox{354 $\pm$ 25 km s$^{-1}$}. These values are consistent within 1$\sigma$ with results reported in the literature \citep{balestra16}.

\begin{table*}
\centering
\begin{threeparttable}
	\caption{Substructures identified in MACS0416 and MACS1206 and their principal features}
 	\label{substructure_features}
        {\small
 		\begin{tabular}{ccccccccc}
		\hline
		\hline
 		Cluster    & Name of Substructure & R.A.\tnote{\emph{a}} & Dec.\tnote{\emph{a}}  & z\tnote{\emph{a}}     & \multicolumn{2}{c}{No. of members\tnote{\emph{b}}} & $\sigma_{sb}$	& $r_{200}$  \\
		           &                &  \multicolumn{2}{c}{(J2000.)}  & &                           &     &  Km s$^{-1}$             & Mpc                 \\ 

(1)    & (2) & (3) & (4)  & (5)     & \multicolumn{2}{c}{(6)}  & (7)	& (8) \\
		\hline
		\hline 
		MACS0416 & MACS0416\_1  & 04 : 15 : 39.3 & -24 : 08 : 50.6 & 0.404 &  6$\pm$4  & (34$\pm$8)   & 306$\pm$64  & 0.612  \\
			 & MACS0416\_2  & 04 : 16 : 11.8 & -24 : 03 : 42.5 & 0.403 & 16$\pm$6  & (78$\pm$12)  & 402$\pm$59  & 0.804  \\
		   	 & MACS0416\_3  & 04 : 16 : 17.2 & -24 : 14 : 58.9 & 0.406 &  3$\pm$3  & (3$\pm$3)    &  -	    & -      \\
			 & MACS0416\_4  & 04 : 15 : 58.7 & -24 : 03 : 07.4 & 0.390 &  4$\pm$3  & (10$\pm$5)   & 203$\pm$58  & 0.409  \\
			 & MACS0416\_5  & 04 : 16 : 20.4 & -24 : 12 : 51.2 & 0.389 & 47$\pm$10 & (89$\pm$13)  & 354$\pm$25  & 0.714  \\
			 & MACS0416\_6  & 04 : 15 : 20.7 & -23 : 54 : 18.0 & 0.399 &  5$\pm$3  & (7$\pm$4)    & 198$\pm$90  & 0.396  \\
			 & MACS0416\_7  & 04 : 16 : 10.5 & -24 : 03 : 45.5 & 0.400 &  9$\pm$4  & (25$\pm$7)   & 172$\pm$55  & 0.344  \\
			 & MACS0416\_8  & 04 : 15 : 25.2 & -23 : 53 : 51.8 & 0.397 &  3$\pm$3  & (6$\pm$4)    & 128$\pm$0   & 0.257  \\
			 & MACS0416\_9  & 04 : 15 : 41.6 & -24 : 08 : 54.1 & 0.397 &  4$\pm$3  & (40$\pm$9)   & 332$\pm$109 & 0.666  \\
			 & MACS0416\_10 & 04 : 15 : 50.3 & -23 : 56 : 22.0 & 0.395 &  9$\pm$4  & (21$\pm$7)   & 197$\pm$68  & 0.396  \\
			 & MACS0416\_11 & 04 : 16 : 13.5 & -23 : 55 : 54.0 & 0.396 &  5$\pm$3  & (21$\pm$7)   & 248$\pm$85  & 0.498  \\
			 & MACS0416\_12 & 04 : 16 : 10.2 & -24 : 04 : 32.9 & 0.396 &  8$\pm$4  & (45$\pm$9)   & 295$\pm$36  & 0.591  \\
			 & MACS0416\_13 & 04 : 16 : 07.3 & -24 : 08 : 43.5 & 0.394 &  5$\pm$3  & (17$\pm$6)   & 250$\pm$57  & 0.503  \\
			 & MACS0416\_14 & 04 : 16 : 21.4 & -24 : 11 : 21.3 & 0.394 &  7$\pm$4  & (77$\pm$12)  & 476$\pm$126 & 0.956  \\
			 & MACS0416\_15 & 04 : 16 : 40.9 & -24 : 04 : 45.5 & 0.395 &  3$\pm$3  & (3$\pm$3)    & -	    & -      \\
		\hline
		MACS1206 & MACS1206\_1  & 12 : 05 : 51.6 & -08 : 41 : 18.4 & 0.440 &  5$\pm$3  & (7$\pm$4)    & 160$\pm$17  & 0.313  \\
			 & MACS1206\_2  & 12 : 06 : 17.8 & -08 : 44 : 01.7 & 0.440 &  3$\pm$3  & (3$\pm$3)    & -	    & -      \\
			 & MACS1206\_3  & 12 : 06 : 31.9 & -08 : 44 : 56.6 & 0.440 &  3$\pm$3  & (6$\pm$4)    & 283$\pm$0   & 0.553  \\
			 & MACS1206\_4  & 12 : 06 : 03.2 & -08 : 48 : 13.5 & 0.440 &  3$\pm$3  & (20$\pm$6)   & 328$\pm$0   & 0.642  \\
			 & MACS1206\_5  & 12 : 06 : 52.5 & -08 : 54 : 12.3 & 0.441 &  5$\pm$3  & (9$\pm$4)    & 119$\pm$41  & 0.232  \\
			 & MACS1206\_6  & 12 : 06 : 20.8 & -08 : 43 : 56.4 & 0.437 &  9$\pm$4  & (10$\pm$5)   & 304$\pm$40  & 0.597  \\
			 & MACS1206\_7  & 12 : 06 : 45.6 & -08 : 37 : 41.4 & 0.438 &  4$\pm$3  & (4$\pm$3)    & -	    & -      \\
			 & MACS1206\_8  & 12 : 05 : 32.4 & -08 : 43 : 08.8 & 0.447 &  3$\pm$3  & (7$\pm$4)    & 239$\pm$0   & 0.465  \\
			 & MACS1206\_9  & 12 : 05 : 58.8 & -08 : 47 : 36.6 & 0.447 &  4$\pm$3  & (16$\pm$6)   & 344$\pm$53  & 0.669  \\
			 & MACS1206\_10 & 12 : 06 : 49.2 & -08 : 54 : 17.6 & 0.445 &  8$\pm$4  & (23$\pm$7)   & 242$\pm$38  & 0.472  \\
			 & MACS1206\_11 & 12 : 06 : 33.2 & -08 : 45 : 22.5 & 0.431 &  5$\pm$3  & (5$\pm$3)    & 125$\pm$111 & 0.245  \\
		\hline
		\hline
		\end{tabular}  
\begin{tablenotes}
\item[\emph{a}]{Central values for each identified substructure.}
\item[\emph{b}]{Number of substructure members  estimated from the spectroscopic sample. In parenthesis we show the number of substructure members estimated from the spectroscopic+photometric sample as described in \S \ref{photometric_substructure_members}.}
\end{tablenotes}
}
\end{threeparttable}
\end{table*}

\subsection{Colour-Magnitude Diagram}\label{cmd_section}

\begin{figure*} 
        \includegraphics[width=0.95\textwidth]{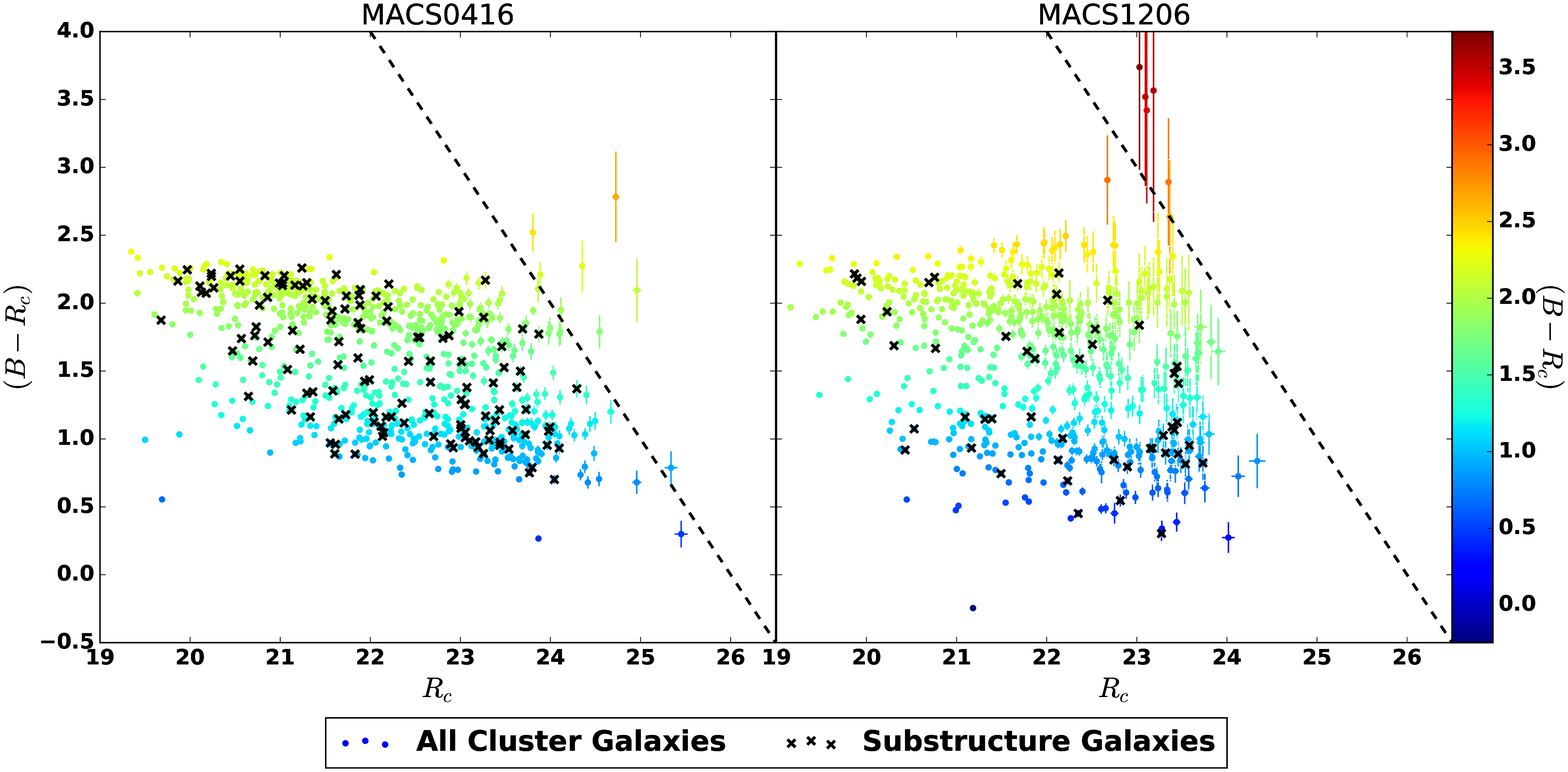}
	\caption{\small{Colour-magnitude diagrams of spectroscopically confirmed cluster members. The colours are measured in the $B$ and $R_{c}$ Subaru bands. Cluster member galaxies are plotted with points whose colours indicate the $(B - R_{c})$ colour and galaxies in substructures are plotted with black crosses. The diagonal dashed lines represent the 3$\sigma$ completeness limit in the $B$ and $R_{c}$ Subaru images. This figure illustrates that galaxies in substructures follow the same colour-magnitude relation as the parent galaxy cluster. Left panel: MACS0416. Right panel: MACS1206.}}
\label{cmd}
\end{figure*}

To analyse the relationship between galaxy colour and environment we study the colour-magnitude properties of galaxies in substructures and in the clusters. For this purpose, using the photometric catalogues from Subaru (see \S \ref{photometric_catalogues}), we performed a match between the spectroscopic members catalogue (see \S \ref{spec_members}) and the photometric catalogue with the objective of obtaining (\mbox{$B - R_{c}$}) colours for the spectroscopic members. The match was performed using the positions in R.A. and Dec. listed for each source in both catalogues. We set an aperture of 1'' as the maximum separation between the matched galaxies. The corresponding (\mbox{$B - R_{c}$}) vs $R_{c}$ colour-magnitude diagram (CMD) is shown in Figure \ref{cmd}.

\begin{figure*} 
        \includegraphics[width=0.47\textwidth]{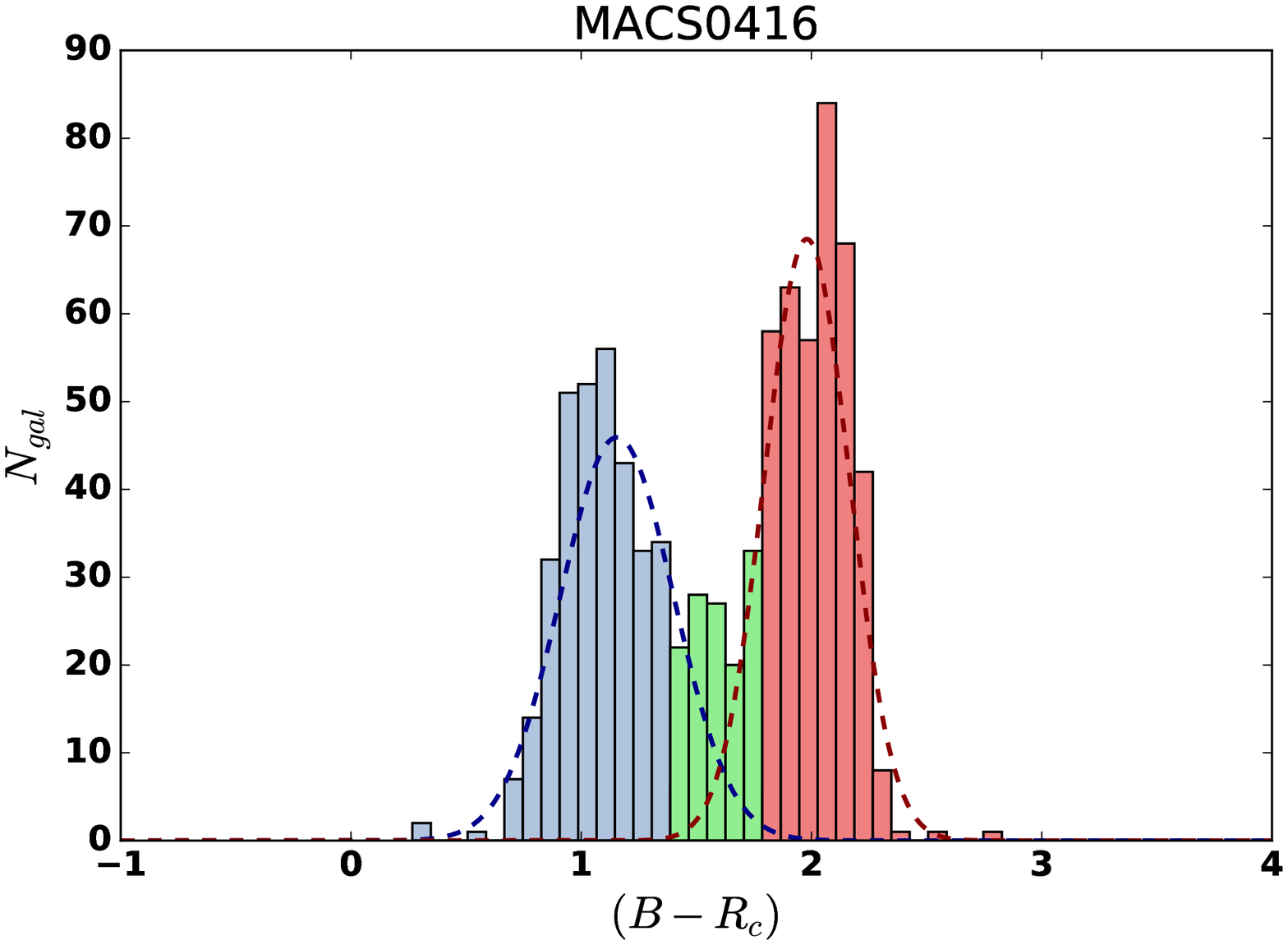}
        \includegraphics[width=0.47\textwidth]{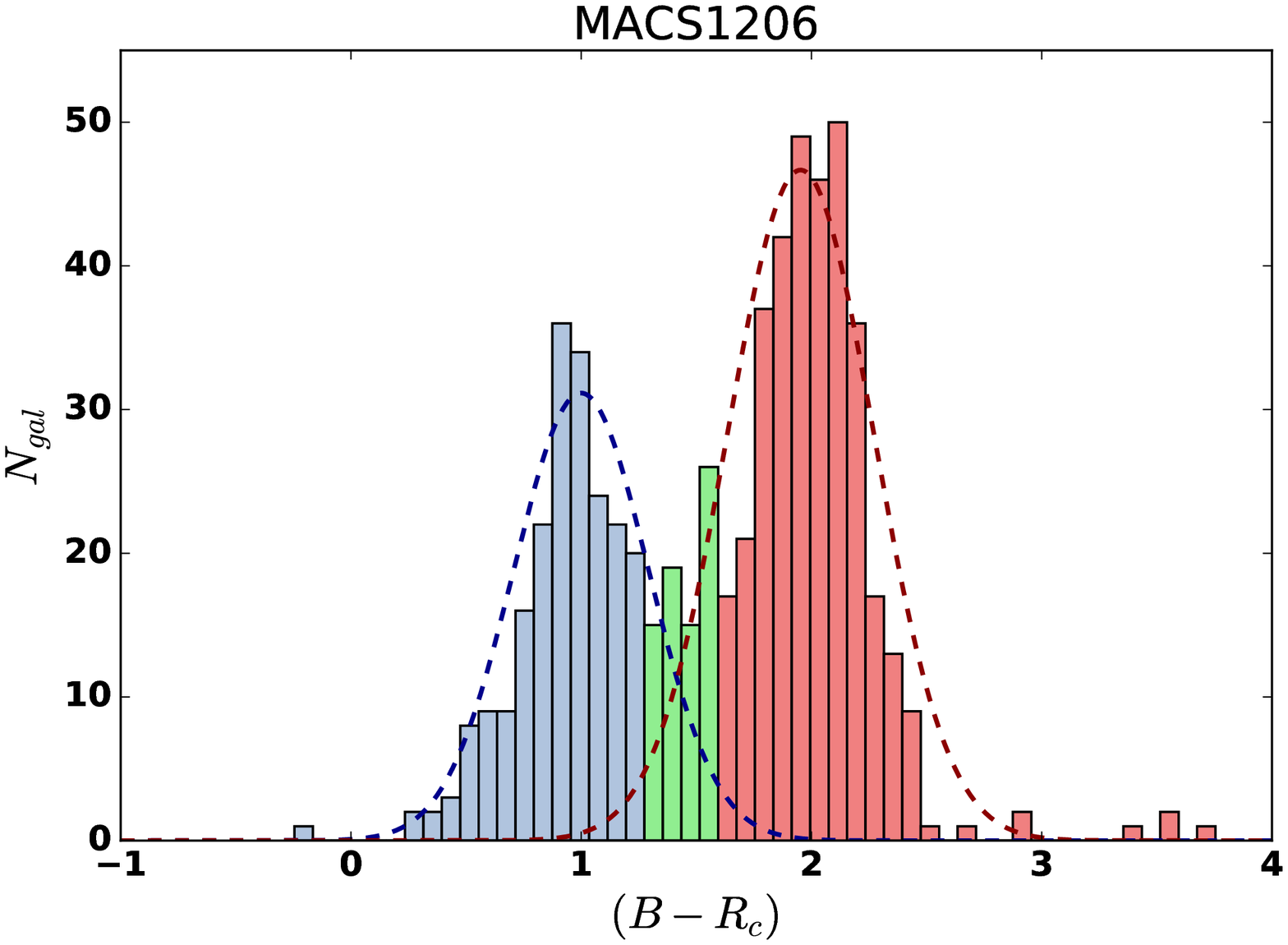}
	\caption{\small{Colour histogram that shows the bimodal distribution in (\mbox{$B - R_{c}$)} colour. Blue and red lines represent the Gaussian fitting of blue and red components determined by GMM. The colour bimodal distribution of galaxies allows us to separate galaxies in blue, red and green according to their colours. Blue galaxies have a colour \mbox{($B - R_{c}$) $\leq$ $\mu_{blue}$ $+$ 1$\sigma_{blue}$}, red galaxies have a colour \mbox{($B - R_{c}$) $\geq$ $\mu_{red}$ $-$ 1$\sigma_{red}$}, and green galaxies are in the \mbox{$\mu_{blue}$ $+$ 1$\sigma_{blue}$ $<$ ($B - R_{c}$) $<$ $\mu_{red}$ $-$ 1$\sigma_{red}$} colour range. Left panel: MACS0416. Right panel: MACS1206.}}
\label{bimodality_plot}
\end{figure*}

We used GMM to fit two Gaussians over the colour distribution to separate galaxy populations (see Figure \ref{bimodality_plot}) in different regions of the CMD. The green valley or transition zone \citep{cortese09} is defined as the space between the blue and red components determined by GMM (e.g \citealt{cortese12}, \citealt{shawinski}). In practice, to separate galaxies according to their colours we fit a Gaussian to the red and blue colour components, separately. Then we estimated the mean ($\mu$) and dispersion ($\sigma$) of each gaussian. We considered as blue those galaxies that have a colour \mbox{($B - R_{c}$) $\leq$ $\mu_{blue}$ $+$ 1$\sigma_{blue}$}, as red those galaxies that have a colour \mbox{($B - R_{c}$) $\geq$ $\mu_{red}$ $-$ 1$\sigma_{red}$}, and as green those galaxies in the \mbox{$\mu_{blue}$ $+$ 1$\sigma_{blue} <$ ($B - R_{c}$) $<$ $\mu_{red}$ $-$ 1$\sigma_{red}$} colour range.

\subsection{Photometric Members}\label{photometric_members} 

As described in \S \ref{completeness_section} in Figure \ref{fig1} we can see that the spectroscopic completeness decreases with the projected distance from the cluster centre, reducing the statistics in the outer regions of the clusters \mbox{($r > r_{200}$)} where most substructures are expected (\citealt{aguerri07}, \citealt{jaffe16}, \citealt{vijayaraghavan}). This affects the characterisation of the properties of infalling galaxies and groups (\citealt{biviano}; \citealt{balestra16}). Thus, in addition to our spectroscopic sample, we built a sample of cluster members selected according to their photometric redshifts (see \S \ref{zphot_section}).

\subsubsection{Cluster Membership}\label{photometric_cluster_members} 

Photometric members were defined as those in the range \mbox{($z_{cl}$ $-$ $\delta_{z}$) $\la$ $z_{phot}$ $\la$ ($z_{cl}$ $+$ $\delta_{z}$)}. To determine $\delta_{z}$ for each cluster we used the available spectroscopic redshifts to calibrate the $z_{phot}$. First, we made a match between the catalogue with spectroscopic members and the catalogue of objects with $z_{phot}$ to estimate the redshift discrepancy \mbox{$\Delta_{z}$ = {\textbar}$z_{spec}$ - $z_{phot}${\textbar}/(1 + $z_{spec}$)}. The match between photometric and spectroscopic catalogues were made using the position in R.A. and Dec. listed for each source in both catalogues. We set an aperture of 1'' as the maximum separation between the matched galaxies. Then we estimated the standard deviation, $\sigma_{\Delta_{z}}$ of the redshift discrepancy distribution to exclude catastrophic identifications, defined here as those with \mbox{$\Delta_{z}$ $\geq$ 3 $\times$ $\sigma_{\Delta_{z}}$}. Finally, we defined $\delta_{z}$ as the standard deviation of the \mbox{{\textbar}$z_{spec}$ - $z_{phot}${\textbar}} distribution in the final clean sample.

Photometric cluster members were defined as those galaxies in the range \mbox{0.359 $< z_{phot} <$ 0.433} and \mbox{0.395 $< z_{phot} <$ 0.483}, for MACS0416 and MACS1206, respectively. Additionally, we used the spectroscopic members to clean the sample of photometrically-selected members from false positives. False positives were defined as those galaxies that were selected as photometric members but whose spectroscopic redshift poses them outside the clusters. In practice, we removed 541 and 275 false positives from the photometrically-selected member catalogues of MACS0416 and MACS1206, respectively. We combined the samples of spectroscopic and photometric members and obtained a sample of 3523 and 2070 \textit{spectrophotometric} members, for MACS0416 and MACS1206, respectively.

\subsubsection{Substructure Membership}\label{photometric_substructure_members} 

Substructure member candidates were defined as those galaxies within $r_{200}$ of each substructure and in the range \mbox{($z_{sub} - \delta_{z}$) $\la$ $z_{phot}$ $\la$ ($z_{sub} + \delta_{z}$)}. The value of $\delta_{z}$ is defined for each substructure in the same way as described in \S \ref{photometric_cluster_members}. For substructures in which we cannot estimate $r_{200}$ we have not added photometric members.

We combined the spectroscopic and photometric sample for substructure, showing the corresponding number of members for each substructure within parenthesis in Column (6) of Table \ref{substructure_features}. Uncertainties in member counts were estimated in the same way as described in \S \ref{ds_substructures}. We note that in some cases (e.g. in MACS1206\_4 and MACS1206\_9) there are substructures that are spatially overlapped and some photometric members are assigned to both substructures. This could be due to the fact that these substructures may be in a merger. This fact is taken into account in \S \ref{photometric_sample} as a source of error due to background contamination.

\subsection{Stellar Masses}\label{sec_stellar_mass}

With only 3 photometric bands available for MACS0416, it is not possible to obtain reliable stellar masses through SED fitting. For this reason we estimated stellar masses in the spectroscopic and spectrophotometric samples with the calibrations of \citet{bell03} for the present-day stellar mass-to-light (M$_{\star}$/L) ratios. These authors used a sample of 12,085 galaxies from the Two Micron All Sky Survey (2MASS, \citealt{skrutskie06}) and the Sloan Digital Sky Survey (SDSS, \citealt{york00}) to study the stellar mass function in the local Universe. To obtain the stellar masses \citet{bell03} used the correlation found by \citet{bell01} (see equation \ref{belleq}) between M$_{\star}$/L and galaxy colours in the rest-frame. \citet{bell03} re-fitted this correlation to find the empirical coefficients (a$_{\lambda}$ and b$_{\lambda}$) that allows the conversion of the SDSS colours into M$_{\star}$/L. The scatter found by \citet{bell03} in the correlation between optical colours and M$_{\star}$/L is $\sim$0.1 dex.

\begin{equation}\label{belleq}
\log_{10}(M_{\star}/L) = a_{\lambda} + (b_{\lambda} \times Colour)
\end{equation}

The relation of \citet{bell03} uses as input the rest-frame photometry. Therefore, to obtain the stellar masses of the galaxies in our sample using Equation \ref{belleq} we must convert our observer-frame photometry to rest-frame photometry. For this, we follow the approach presented in the {\sc appendix} B of \citet{mei09} to convert the \mbox{$(B - R)$} observer-frame colour to rest-frame SDSS colour and the $B$ observer-frame magnitude to rest-frame SDSS magnitude. \citet{mei09} found that there is a correlation between observer-frame photometry and rest frame photometry given by

\begin{equation}\label{rf_color}
C_{rf} = A + B \times C_{obs}   
\end{equation}

\begin{equation}\label{rf_magnitude}   
M_{rf} - m_{obs} = \alpha + \beta \times C_{obs} 
\end{equation}

Zero-points ($A$ and $\alpha$) and slopes ($B$ and $\beta$) in equations \ref{rf_color} and \ref{rf_magnitude} were estimated using a set of synthetic stellar population models from \citeauthor{bruzual} (\citeyear{bruzual}; BC03) with ten different metallicities from 0.4 to 2.5$Z_{\odot}$, two laws of star formation rate (instantaneous burst and exponentially decaying with e-folding time \mbox{$\tau$ = 1 Gyr}) and \citet{salpeter95} IMF. The BC03 spectral library spans the range in formation redshift \mbox{2.0 $<$ z$_{f}$ $<$ 5.0}. The models were generated with \texttt{EzGal}\footnote{\url{http://www.baryons.org/ezgal/}} \citep{mancone12} and for each one we extracted the observed and rest-frame colours and magnitudes at the redshift of each cluster and fitted the linear relation. Zero-points and slopes were derived following the approach presented by \citet{dang06}. In summary we fit straight lines to all the possible couples of points in the colour-colour and magnitude-colour planes and obtain the distributions of slopes and intercepts. Zero-points and slopes are defined as the median of each distribution, while their uncertainties are taken as the 1$\sigma$ width of the distributions. 

In practice, stellar masses (M$_{\star}$) of the galaxies in our sample were obtained converting the \mbox{$(B - R)$} observer-frame colour to \mbox{(\textit{u'} - \textit{g'})} rest-frame colour and the $B$ observer-frame magnitude to rest-frame absolute $M_{g'}$ magnitude. The values of a$_{g}$ and b$_{g}$ were taken from Table A7 presented by \citet{bell03}. Stellar mass uncertainties were estimated using a technique based on a Monte Carlo approach. The observer-frame colour, observer-frame magnitudes, zero-points ($A$ and $\alpha$), slopes ($B$ and $\beta$), rest-frame colours and rest-frame magnitudes were perturbed by a random value, $\epsilon$, in the range \mbox{-$\Delta$ $<$ $\epsilon$ $<$ +$\Delta$}, where $\Delta$ corresponds to the error in each variable. We generate 20 simulated objects for each real object. Finally, we estimated the uncertainties on stellar mass evaluating the 1$\sigma$ asymmetric width of the simulated stellar mass distribution. The stellar masses obtained are in the range \mbox{8 $ \leq \log(M_{\star}/M_{\odot}) \leq$ 11.5}  for both MACS0416 and MACS1206. The uncertainties in the stellar masses are in the range \mbox{0.1 - 0.2dex}.

\subsection{Colour Fractions} 

Galaxy colours are related to internal properties. For example, red galaxies are more likely passive or quiescent, without star-formation. Blue galaxies, on the contrary, are more likely star forming. For this reason, we are interested in examining the relationship between galaxy colours and the local environment. We estimated the fraction of blue, green and red galaxies as a function of projected distance from the cluster centre or substructure centre, as a function of $R_{c}$ magnitude, and as a function of stellar mass ($M_{\star}$). Blue, green and red populations of galaxies were defined using the colour bimodality described in \S \ref{cmd_section}. The colour fraction in the clusters were estimated in the sample of cluster galaxies that are not part of a substructure.

\subsubsection{Spectroscopic Sample}\label{spectroscopic_fractions}

The colour fractions of galaxies in the spectroscopic sample were estimated taking into account the effect of spectroscopic incompleteness (see \S \ref{completeness_section}). In practice, we corrected galaxy counts by weighting each galaxy with \mbox{$W_{i} = 1/C$}, where $C$ is the spectroscopic completeness estimated in bins of distance and magnitude as described in \S \ref{completeness_section}. To increase the statistics in the spectroscopic sample, we merged the spectroscopic member catalogues for both clusters. We also merged the catalogue for all identified substructures. We estimated the fractions of blue (\textit{f$_{b}$}), green (\textit{f$_{g}$}) and red (\textit{f$_{r}$}) galaxies in the whole cluster sample and in the substructures. Error-bars on the fractions were computed using a method based on statistical inference in absence of background contamination (\citealt{dagostini}, \citealt{cameron}). In practice, the uncertainties on the spectroscopic colour fractions were defined as the width of the binomial 68\% confidence intervals as discussed in \citet{cameron}.

The optically selected red sequence may be contaminated by dust-obscured star-forming galaxies (see e.g.: \citealt{wolf05}, \citealt{haines08}). For this reason our colour fractions are corrected for contamination by dusty star-forming galaxies as illustrated in \mbox{Section \ref{photometric_sample}}.

\begin{figure*} 
        \includegraphics[width=1.\textwidth]{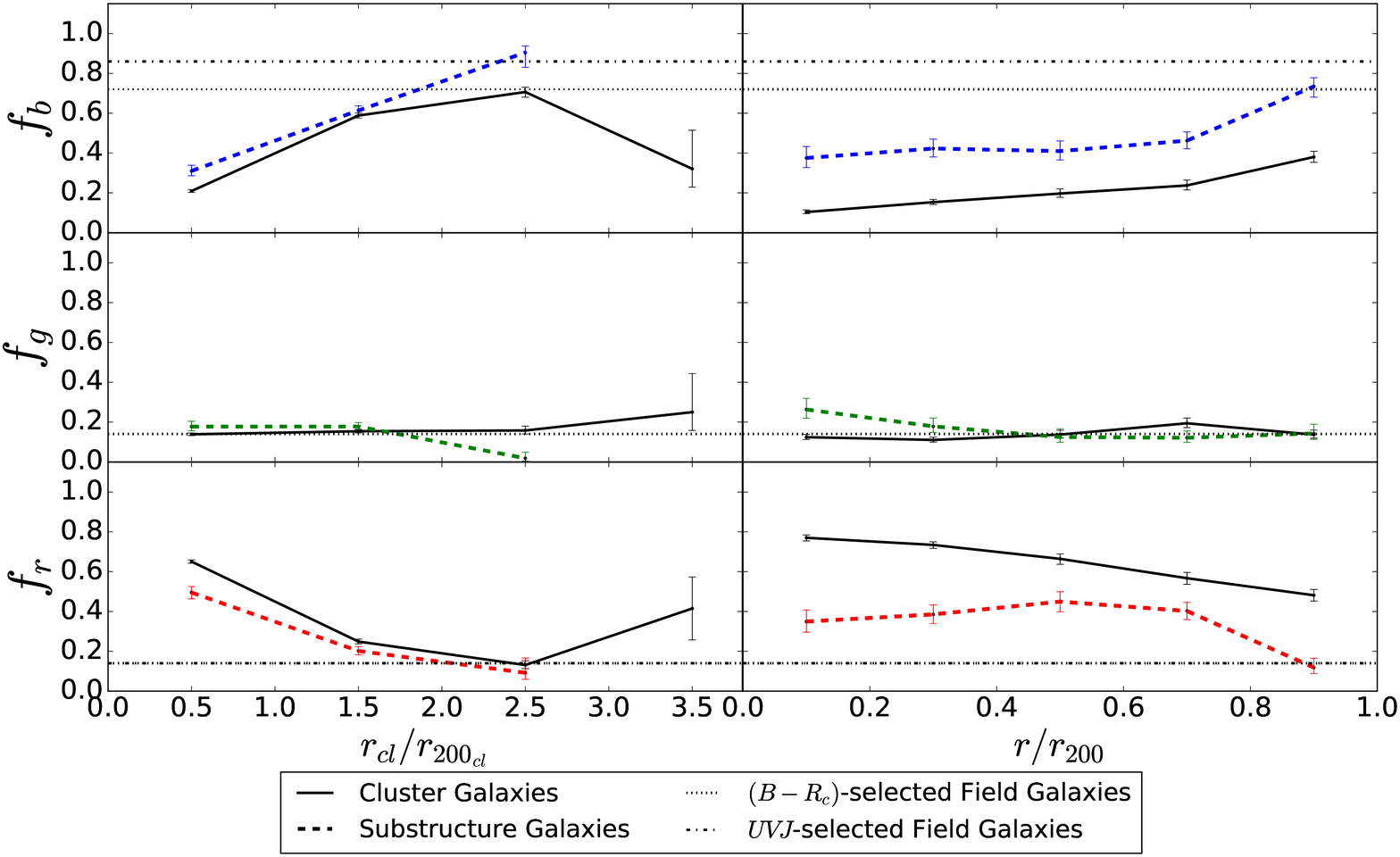} 
	\caption{\small{Left panel: Fraction of blue ($f_{b}$), green ($f_{g}$) and red ($f_{r}$) galaxies as a function of projected distance from the cluster centre. Right panel: $f_{b}$, $f_{g}$ and $f_{r}$ as a function of projected distance from cluster or substructure centre. The dotted horizontal black lines represent $f_{b}$, $f_{g}$ and $f_{r}$ in the field. The dash-dotted horizontal black lines represent the fractions of star-forming ($f_{SF}$) and passive galaxies ($f_{P}$) in the field. Galaxy fractions in the field correspond to a mean value. Galaxy clusters and substructures have lower fractions of blue galaxies than the fraction of star-forming galaxies observed in the field, indicating that cluster and substructures are more efficient in producing red galaxies.}}
\label{fig5}
\end{figure*}

\begin{figure*} 
        \includegraphics[width=1.\textwidth]{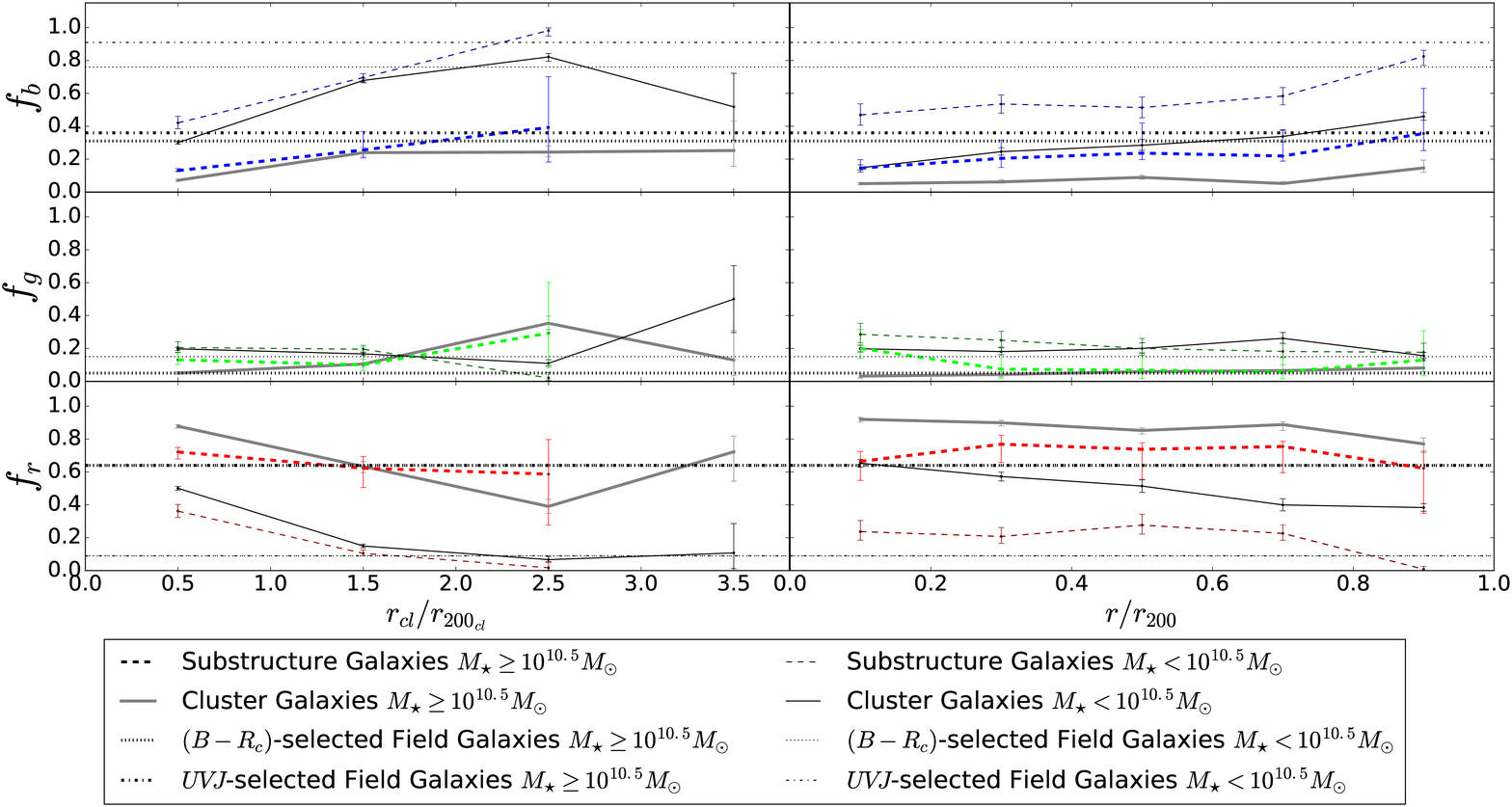}
	\caption{\small{Left panel: Fraction of blue ($f_{b}$), green ($f_{g}$) and red ($f_{r}$) galaxies as a function of projected distance from the cluster centre. Right panel: $f_{b}$, $f_{g}$ and $f_{r}$ as a function of projected distance from cluster or substructure centre. The dotted horizontal black lines represent $f_{b}$, $f_{g}$ and $f_{r}$ in the field. The dash-dotted horizontal black lines represent the fractions of star-forming ($f_{SF}$) and passive galaxies ($f_{P}$) in the field. Galaxy fractions as a function of projected distance in all environments were estimated by separating galaxies in two bins of stellar mass in massive \mbox{($M_{\star} \geq 10^{10.5}M_{\odot}$)} and less massive galaxies \mbox{($M_{\star} < 10^{10.5}M_{\odot}$)}. The fraction of red galaxies is higher for massive galaxies independent of the environment. However, in dense environments the fraction of red galaxies is higher than in the field, indicating that cluster and substructures are more efficient in producing red galaxies.}}
\label{fig6}
\end{figure*}

\begin{figure*} 
        \includegraphics[width=0.95\textwidth]{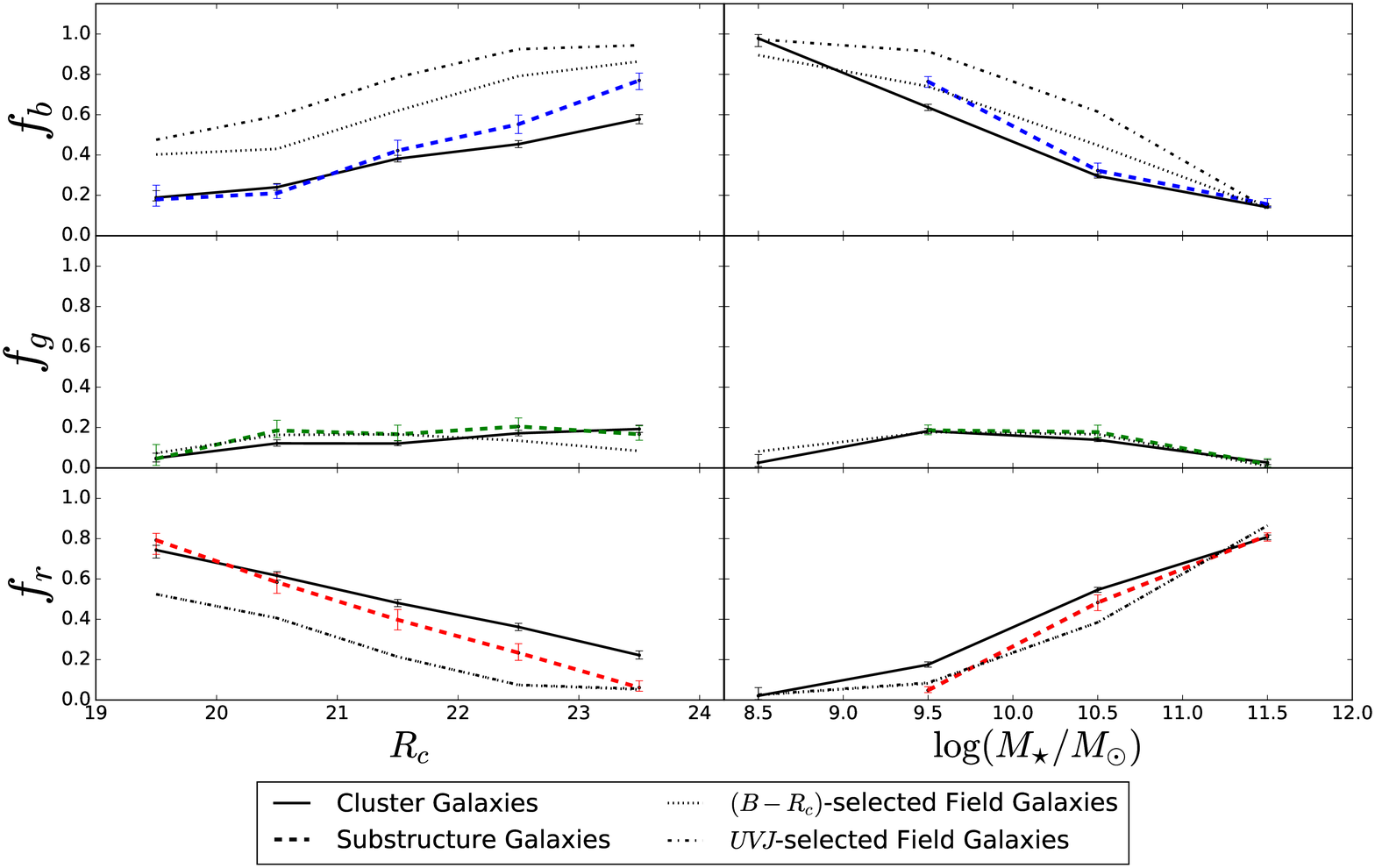}
	\caption{\small{Left panel: Fractions of blue ($f_{b}$), green ($f_{g}$) and red ($f_{r}$) galaxies in galaxy clusters and substructures as a function of $R_{c}$ apparent magnitude. Right panel: \textit{f$_{b}$}, \textit{f$_{g}$} and \textit{f$_{r}$} as a function of $\log(M_{\star}/M_{\odot})$. The dash-dotted black lines represent the fractions of star-forming ($f_{SF}$) and passive galaxies ($f_{P}$) in the field. The fraction of blue galaxies increases with increasing $R_{c}$ magnitude and decreasing stellar mass in all environments, but the fraction of star-forming galaxies in the field is higher than the fraction of blue galaxies in cluster or substructures.}}
\label{fig7}
\end{figure*}

Our results are shown in Figures \ref{fig5}, \ref{fig6} and \ref{fig7}, in which we present \textit{f$_{b}$}, \textit{f$_{g}$} and \textit{f$_{r}$} for the clusters and substructures. The left panel of Figures \ref{fig5} and \ref{fig6} shows \textit{f$_{b}$}, \textit{f$_{g}$} and \textit{f$_{r}$} as a function of projected distance from the cluster centre normalised by $r_{200_{cl}}$. The right panel presents \textit{f$_{b}$}, \textit{f$_{g}$} and \textit{f$_{r}$} as a function of projected distance from the cluster centre and substructure centre, for cluster and substructure galaxies, respectively. These projected distances were normalised by $r_{200}$ using the values for the galaxy clusters or substructures accordingly. Instead, the left panel of Figure \ref{fig7} shows \textit{f$_{b}$}, \textit{f$_{g}$} and \textit{f$_{r}$} as a function of $R_{c}$ magnitude for cluster and substructure galaxies, and the right panel presents \textit{f$_{b}$}, \textit{f$_{g}$} and \textit{f$_{r}$} as a function of $\log(M_{\star}/M_{\odot})$.

In Figures \ref{fig5}, \ref{fig6} and \ref{fig7} we have added the observed fractions of field galaxies for comparison purposes. The dotted line represents the fraction of blue, green and red field galaxies. These fractions were estimated using the colour bimodality, as for the spectroscopic sample. The counts in the red and blue fractions were corrected by the contamination in the red sequence due to dusty star-forming galaxies. The dash-dotted line represents the fraction of star-forming ($f_{SF}$) and passive ($f_{P}$) galaxies in the field, accordingly. The $f_{SF}$ and $f_{P}$ were estimated using the $UVJ$-diagram (\citealt{wuyts07}, \citealt{patel11}, \citealt{nantais16}). The selection of the field sample is described in \S \ref{photometric_sample}. These results will be discussed in \S \ref{discussion}.

\subsubsection{Spectrophotometric Sample}\label{photometric_sample} 

We determine the number of blue, green and red galaxies, in the spectrophotometric sample, using the same approach as for the spectroscopic sample (see \S \ref{cmd_section}). However, when using photometric redshifts to select cluster members, we are likely to include field interlopers as a result of the uncertainties on $z_{phot}$, which are larger than those of spectroscopic redshifts. Nevertheless, we can statistically estimate the expected contribution of field interlopers at the cluster redshift and correct our colour fractions by the field contamination. 

To estimate the contribution of field galaxies in our estimated colour fractions we use the COSMOS/UltraVISTA catalogue published by \citet{muzzin13} to build a subsample of 3,166 field galaxies at \mbox{$0.36 < z < 0.46$}. The COSMOS/UltraVISTA field sample covers \mbox{1.62 deg$^{2}$} of the Cosmic Evolution Survey (COSMOS, \citealt{scoville07}) and the sources in the catalogue have been selected from the UltraVISTA K$_{s}$ survey \citep{mccracken12} reaching a depth of \mbox{K$_{s,tot}$ = 23.4 mag} (90\% completeness). The UltraVISTA survey was carried out with the VISTA InfraRed CAMera (VIRCAM, \citealt{dalton06}) on the Visible and Infrared Survey Telescope for Astronomy (VISTA, \citealt{emerson06}) at the ESO/Paranal Observatory. The COSMOS/UltraVISTA sample is available in the UltraVISTA web repository\footnote{\label{note7}\url{http://www.strw.leidenuniv.nl/galaxyevolution/ULTRAVISTA/Ultravista/K-selected.html}} and provides a point-spread function (PSF) matched photometry in 30 photometric bands covering the wavelength range of \mbox{0.15 - 24 $\mu$m}, $z_{phot}$, stellar masses and rest-frame $U$, $V$, and $J$ photometry of 262,615 sources, as discussed in \citet{muzzin13}. 

The blue, green and red populations of our sample of 3,166 field galaxies at \mbox{$0.36 < z < 0.46$} were defined in the same way as it was done with the clusters, using the colour bimodality described in \S \ref{cmd_section}. The colour fractions (\textit{f$_{b}$}, \textit{f$_{g}$} and \textit{f$_{r}$}) of field galaxies were used to infer the contamination in the colour fractions in clusters/substructures due to field background and foreground sources. To correct our colour fractions in the spectrophotometric sample for background contamination we use an approach based on statistical inference. In practice we build a posterior probability distribution in which the colour fractions from clusters/substructures are considered as signal and the colour fractions from field are considered as noise. The full mathematical derivation of this approach is presented in \citet{dagostini} (see also \citealt{cameron}). In the same way as \citet{cerulo17} we defined our background-corrected colour fractions in galaxy clusters/substructures as the median of the posterior probability distribution. The uncertainty in our background corrected colour fractions was defined as the 68\% credible interval of the distribution. 

Our blue and red galaxy fractions in cluster, substructures and the field were further corrected for contamination of dusty star-forming galaxies on the red sequence using the $UVJ$-diagram as a diagnostic tool to distinguish young, star-forming and dusty from old and passive galaxies (\citealt{wuyts07}, \citealt{patel11}, \citealt{nantais16}). We used the $UVJ$ selection proposed by \citet{williams09} to select quiescent galaxies and clean the red sequence in the clusters, substructures and field from dusty star-forming contaminants. In our sample of 3,166 field galaxies at \mbox{$0.36 < z < 0.46$} from COSMOS/UltraVISTA we found that 40\% of the galaxies selected on the field red sequence are dusty-star forming galaxies. To estimate the contamination of dusty star-forming galaxies in the red sequence in the cluster centres \mbox($r < r_{200}$) we matched, using R.A. and Dec., the spectrophotometric member catalogue with the photometric catalogue from HST. We set an aperture of 1" as the maximum separation between the matched galaxies. We found that 5\% of the selected galaxies in the red sequence of the clusters in the inner regions \mbox($r < r_{200}$) are dusty star-forming galaxies. Finally, to estimate the contamination in the substructures and in outer regions of the clusters \mbox($r > r_{200}$) we used the groups catalogue from COSMOS published by \citet{george11} limited to the redshift of the clusters (\mbox{$0.36 < z < 0.46$}). The COSMOS groups in this redshift range are similar, in terms of mass and size, to the selected substructures. We found that 17\% of the galaxies selected on the red sequence of the COSMOS groups are dusty star-forming. Our results will be discussed in \S \ref{discussion}. 

\begin{figure*}
        \includegraphics[width=1\textwidth]{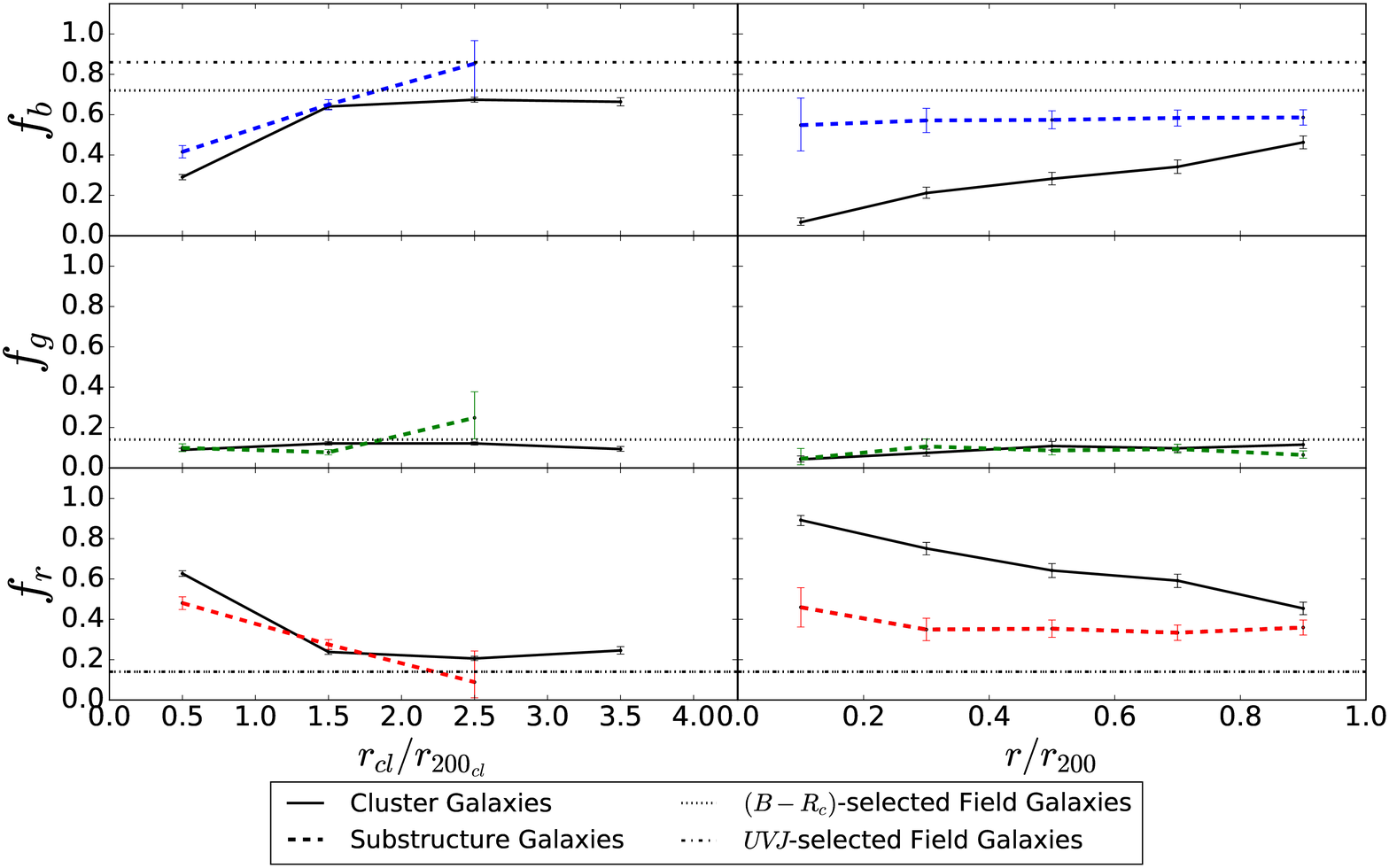}
	\caption{\small{Background-corrected colour fractions for clusters and substructures as a function of projected distance normalised by $r_{200}$. Left panel: Fractions of blue (\textit{f$_{b}$}), green (\textit{f$_{g}$}) and red (\textit{f$_{r}$}) galaxies as a function of projected distance from the cluster centre. Right panel: \textit{f$_{b}$}, \textit{f$_{g}$} and \textit{f$_{r}$} as a function of projected distance from cluster or substructure centre, accordingly. The dotted horizontal black line represents the \textit{f$_{b}$}, \textit{f$_{g}$} and \textit{f$_{r}$} of field galaxies, while the dash-dotted horizontal black lines represent the fractions of star-forming ($f_{SF}$) and passive galaxies ($f_{P}$) in the field. Galaxy fractions in the field are mean values. In general, galaxies in overdensities (clusters and substructures) have higher \textit{f$_{r}$} than in the field.}}
\label{fig8}
\end{figure*}

\begin{figure*}
        \includegraphics[width=1\textwidth]{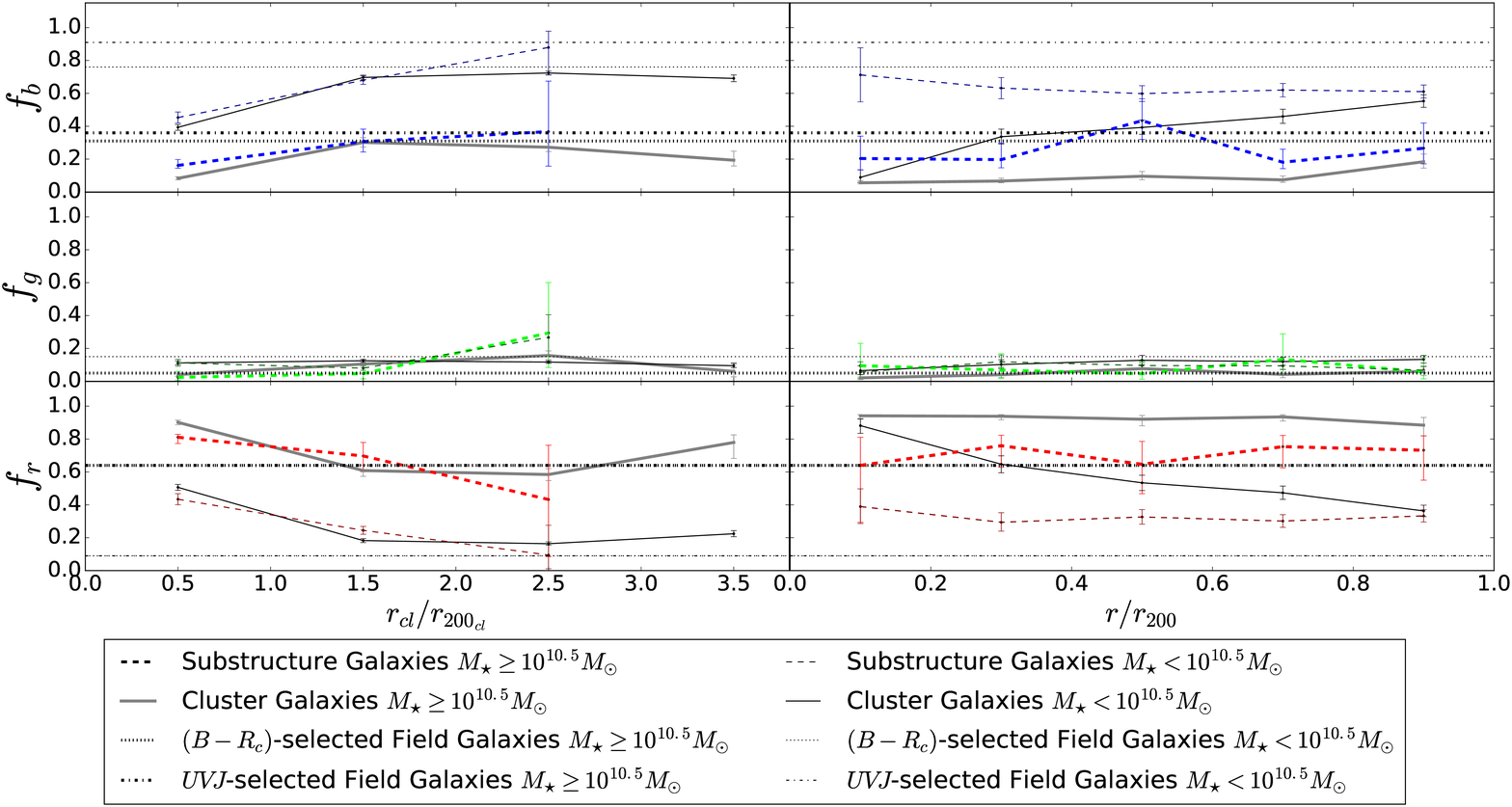}
	\caption{\small{Background-corrected colour fractions for clusters and substructures as a function of projected distance normalised by $r_{200}$. Left panel: Fractions of blue (\textit{f$_{b}$}), green (\textit{f$_{g}$}) and red (\textit{f$_{r}$}) galaxies as a function of projected distance from the cluster centre. Right panel: \textit{f$_{b}$}, \textit{f$_{g}$} and \textit{f$_{r}$} as a function of projected distance from cluster or substructure centre, accordingly. The dotted horizontal black line represents the \textit{f$_{b}$}, \textit{f$_{g}$} and \textit{f$_{r}$} of field galaxies, while the dash-dotted horizontal black lines represent the fractions of star-forming ($f_{SF}$) and passive galaxies ($f_{P}$) in the field. Galaxy fractions in the field are mean values. Galaxy fractions as a function of projected distance in all environments were estimated by separating galaxies in two bins of stellar mass in massive \mbox{($M_{\star} \geq 10^{10.5}M_{\odot}$)} and less massive galaxies \mbox{($M_{\star} < 10^{10.5}M_{\odot}$)}. In general, independent of the stellar mass, galaxies in dense environments such as clusters and substructures tend to be redder than galaxies in the field.}}
\label{fig9}
\end{figure*}

\begin{figure*} 
        \includegraphics[width=0.95\textwidth]{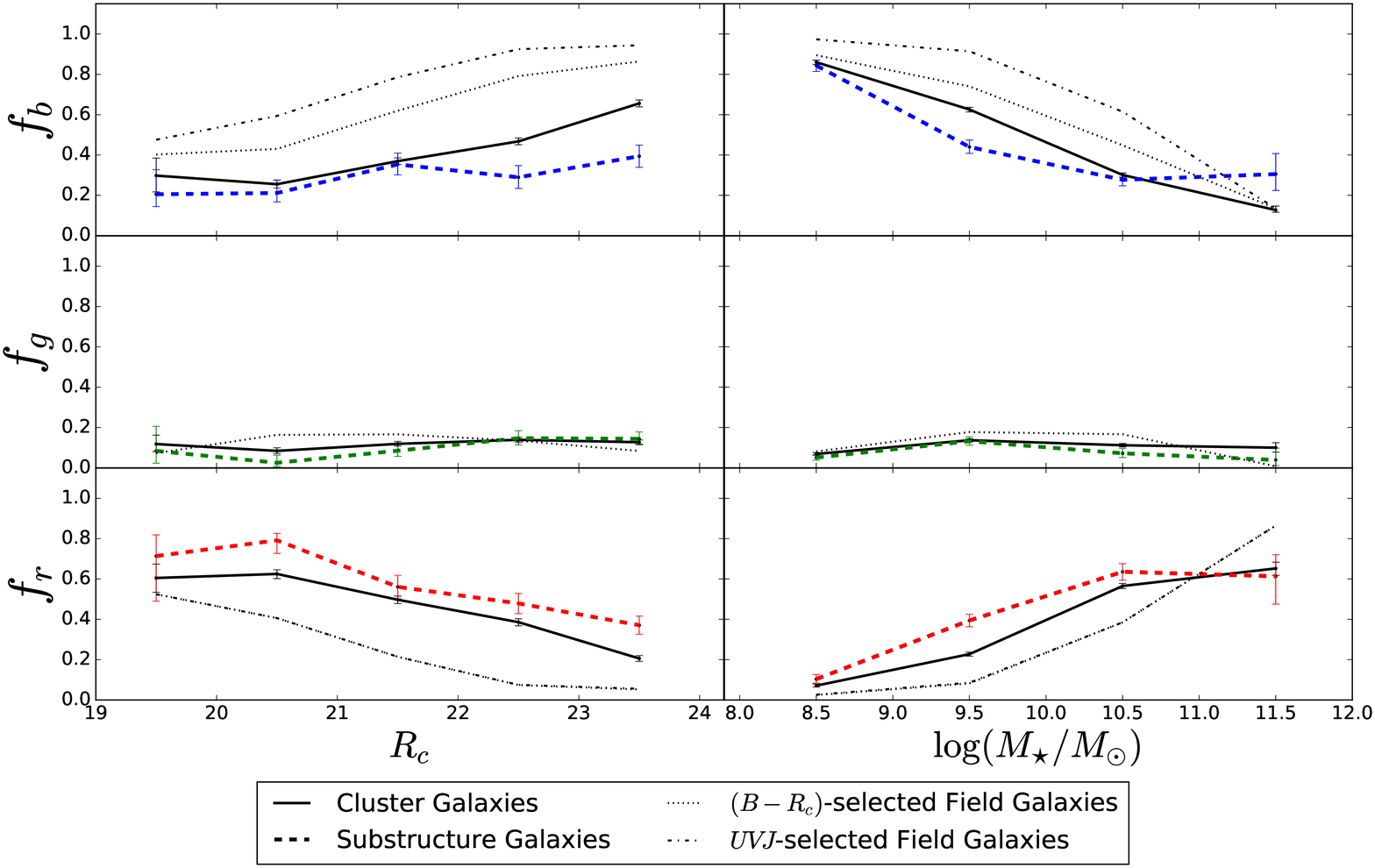}
	\caption{\small{Left panel: Background-corrected colour fractions for clusters and substructures as a function of $R_{c}$ apparent magnitude.
Right panel: Background-corrected colour fractions for clusters and substructures as a function of $\log(M_{\star}/M_{\odot})$. The dashed black curve represents the \textit{f$_{b}$}, \textit{f$_{g}$} and \textit{f$_{r}$} of field galaxies, while the dash-dotted black lines represent the fractions of star-forming ($f_{SF}$) and passive galaxies ($f_{P}$) in the field. Galaxy fractions in the field were estimated in bins of $R_{c}$-band magnitude or stellar mass. The fraction of red or quiescent galaxies increases towards bright magnitudes and massive galaxies in all environments.}}
\label{fig10}
\end{figure*}

Figures \ref{fig8}, \ref{fig9} and \ref{fig10} show the \textit{f$_{b}$}, \textit{f$_{g}$} and \textit{f$_{r}$} as a function of distance from the overdensity centre, $R_{c}$ magnitude and stellar mass for cluster and substructure galaxies. These results will be discussed in \S \ref{discussion}. In the figures we have added the fractions of blue, green and red field galaxies. These fractions were estimated in our sample of 3,166 field galaxies at \mbox{$0.36 < z < 0.46$} after removing group members according to the catalogue of \citet{george11}.

Since we found that 40\% of the galaxies selected on the field red sequence are dusty-star forming galaxies, we use the $UVJ$-diagram (\citealt{wuyts07}, \citealt{patel11}, \citealt{nantais16}) to estimate the fraction of star-forming ($f_{SF}$) and passive ($f_{P}$) galaxies in the field. In the clusters/substructures we cannot use the $UVJ$-diagram to estimate the $f_{SF}$ and $f_{P}$ because there are not data in the $UVJ$-bands that cover the outer regions of the clusters \mbox($r > r_{200}$). However, we correct the colour fractions for cluster and substructure members, in both the spectroscopic and spectrophotometric samples, for the contamination from dusty star-forming galaxies derived from the CLASH HST data and the COSMOS groups. The fractions shown in Figures \ref{fig5} -- \ref{fig10} are all corrected for contamination from dusty star-forming galaxies.

\subsection{Substructure Quenching Efficiency}\label{eq_section} 

\begin{figure*} 
        \includegraphics[width=0.95\textwidth]{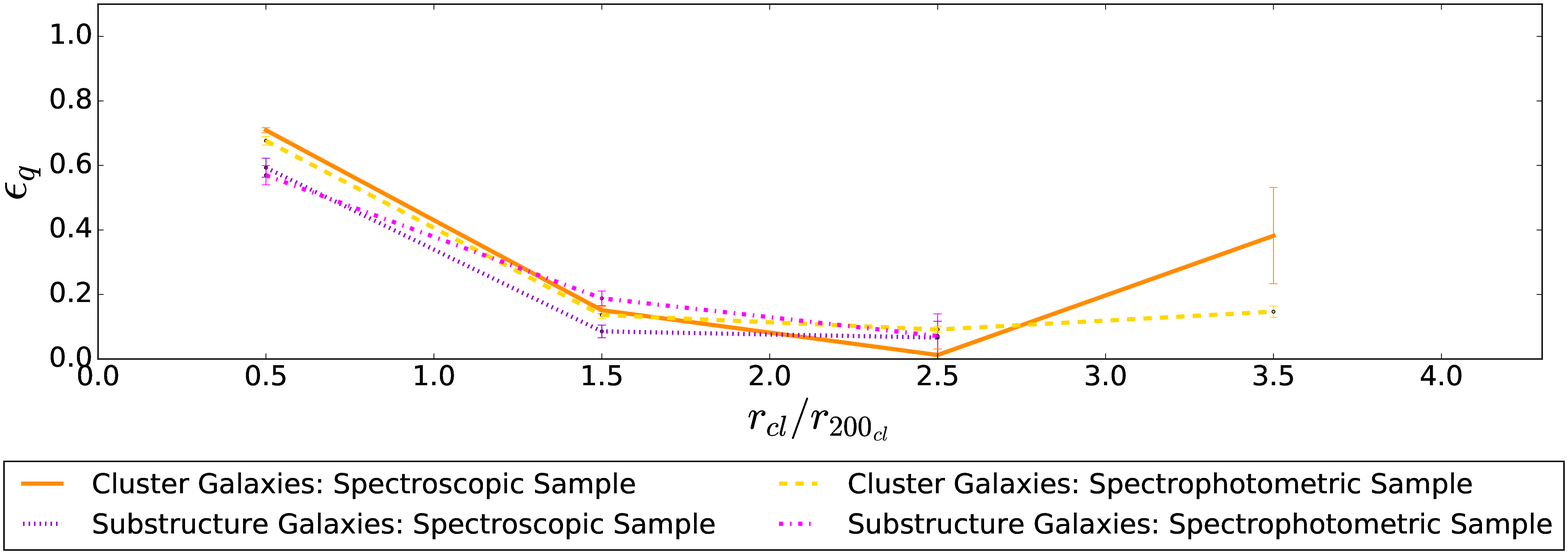}
	\caption{\small{Environmental quenching efficiency in clusters and substructures calculated with equation \ref{quenching_efficiency} \citep{peng10} as a function of projected distance from the cluster centre normalised by $r_{200_{cl}}$. The solid brown and dashed yellow lines represent the $\epsilon_{q}$ in the clusters for the spectroscopic (see \S \ref{spec_members}) and spectrophotometric (see \S \ref{photometric_members}) samples, respectively. The dotted indigo and dash-dotted magenta lines represent the $\epsilon_{q}$ in the substructures for the spectroscopic and spectrophotometric samples, respectively. In the outskirts of clusters \mbox{($r > r_{200_{cl}}$)} the $\epsilon_{q}$ of substructures becomes comparable to the $\epsilon_{q}$ of clusters.}}
\label{fig11}
\end{figure*}

In our analysis, we assume that the colour change of galaxies in clusters/substructures is associated with the migration of galaxies from the blue cloud to the red sequence in the CMD. This migration is caused by the quenching of star-formation (e.g \citealt{peng10}). To quantify the role of the substructures in the star formation quenching, or pre-processing, we estimated the environmental quenching efficiency ($\epsilon_{q}$; \citealt{peng10}, \citealt{peng12}, \citealt{nantais16}), or conversion fraction, because it quantifies the fraction of galaxies that would be blue in the field but are red in a denser environment (e.g cluster or substructures; \citealt{peng10}). The environmental quenching efficiency is defined as follows:

\begin{equation}\label{quenching_efficiency}
\epsilon_{q} = \frac{(f_{r,dense} - f_{r,field})}{f_{b,field}}~,
\end{equation}
where in this equation, $\epsilon_{q}$ is the environmental quenching efficiency, $f_{r,dense}$ is the red fraction in a dense environment (cluster or substructure), and $f_{r,field}$ and $f_{b,field}$ are the red and blue fractions in the field, respectively. In our sample, we estimated $\epsilon_{q}$ in clusters and substructures as a function of projected distance from the cluster centre, normalised by $r_{200}$. Besides, we estimated the $\epsilon_{q}$ in cluster and substructure as a function of projected distance from cluster or substructure centre normalised by $r_{200}$ of the cluster or substructure, accordingly. $\epsilon_{q}$ is calculated in the same bins used to derive colour fractions. The uncertainty on the environmental quenching efficiency was defined, in each bin of projected distance, as the 68\% width of the distribution of $\epsilon_{q}$ after 100,000 Monte-Carlo iterations performed by randomly perturbing the values of the colour fractions within their error bars. In Table \ref{eq_table} we summarise our mean environmental quenching efficiency in clusters and substructures, estimated using the spectroscopic and spectrophotometric samples. Our results are presented in Figures \ref{fig11} and \ref{fig13}, and will be discussed in \S \ref{discussion}.

\begin{figure*} 
        \includegraphics[width=0.95\textwidth]{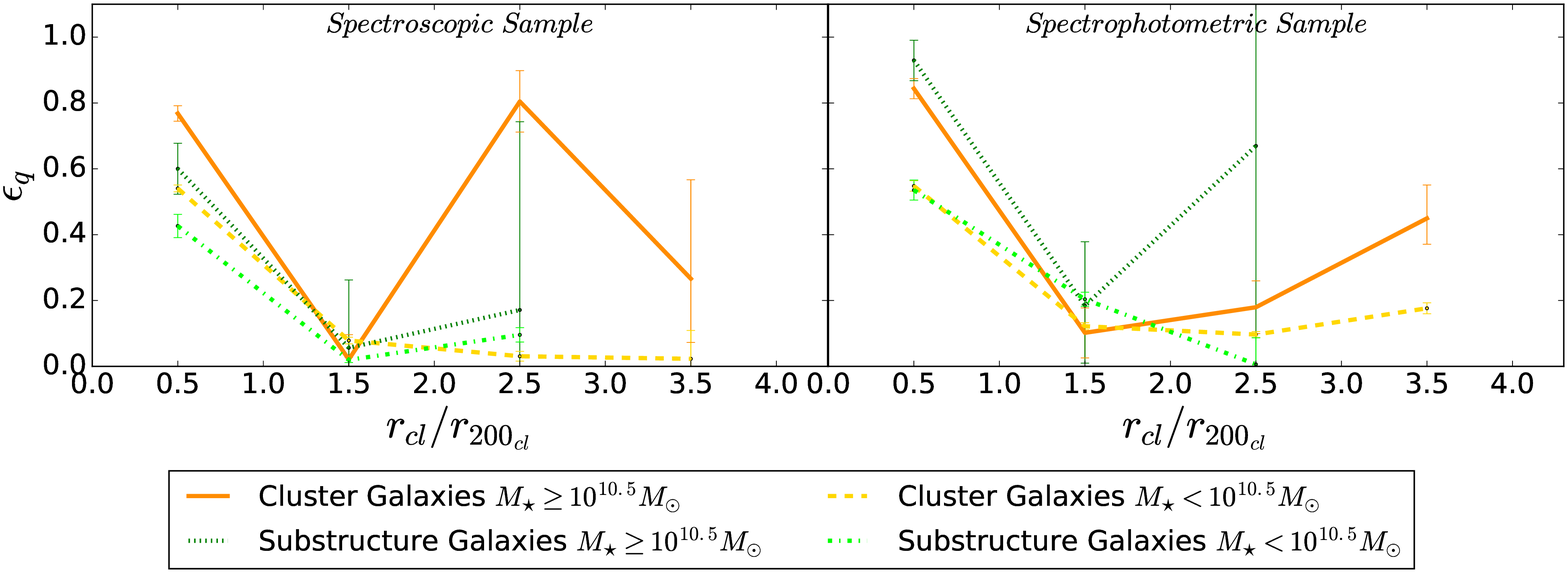}
	\caption{\small{Environmental quenching efficiency separated by mass in massive \mbox{($M_{\star} \geq 10^{10.5}M_{\odot}$)} and less massive galaxies \mbox{($M_{\star} < 10^{10.5}M_{\odot}$)} in clusters and substructures calculated with equation \ref{quenching_efficiency} \citep{peng10} as a function of projected distance from the cluster centre normalised by $r_{200_{cl}}$. The solid orange and dashed gold lines represent the $\epsilon_{q}$ in the clusters for massive and less massive galaxies, respectively. The green dotted and dash-dotted lines represent the $\epsilon_{q}$ in substructures. Left panel: spectroscopic sample. Right panel: spectrophotometric sample. The environmental quenching efficiency is higher in massive galaxies.}}
\label{fig12}
\end{figure*}

\begin{figure*} 
        \includegraphics[width=0.95\textwidth]{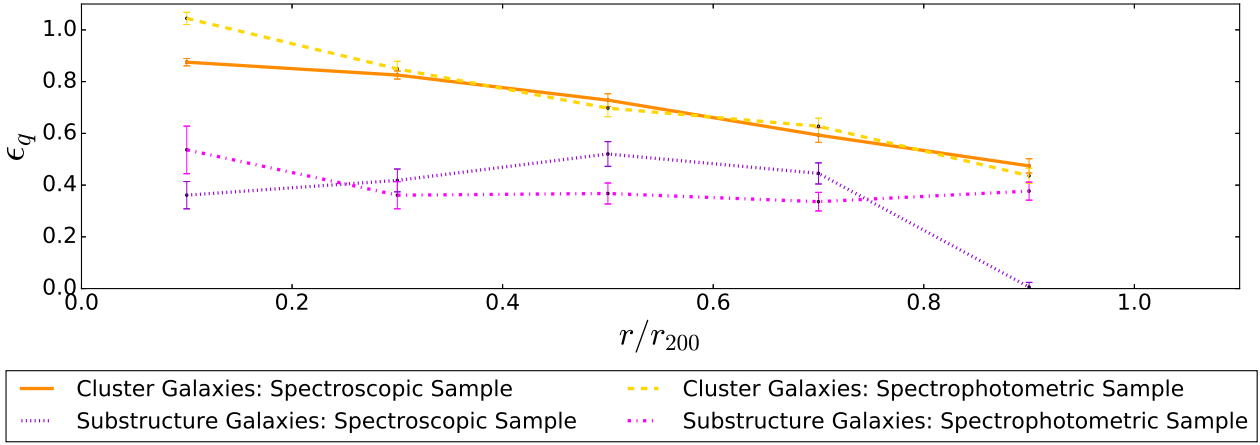}
	\caption{\small{Environmental quenching efficiency in clusters and substructures calculated with equation \ref{quenching_efficiency} \citep{peng10} as a function of projected distance from the cluster centre or substructure centre normalised by $r_{200}$ of the cluster or substructure, accordingly. The solid brown and dashed yellow lines represent the $\epsilon_{q}$ in the clusters for the spectroscopic (see \S \ref{spec_members}) and spectrophotometric (see \S \ref{photometric_members}) samples, respectively. The dotted indigo and dash-dotted magenta lines represent the $\epsilon_{q}$ in the substructures for the spectroscopic and spectrophotometric samples, respectively. In the innermost regions \mbox{($r < r_{200}$)} of the overdensity (either cluster or substructure) clusters are more efficient than substructures in quenching star formation.}}
\label{fig13}
\end{figure*}

\begin{figure*} 
        \includegraphics[width=0.95\textwidth]{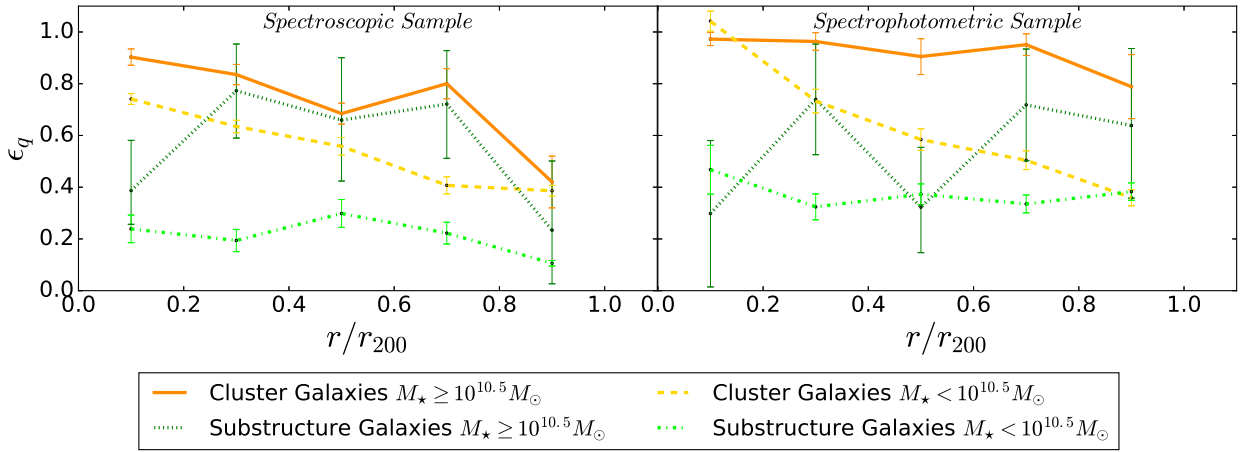}
	\caption{\small{Environmental quenching efficiency separated by mass in massive \mbox{($M_{\star} \geq 10^{10.5}M_{\odot}$)} and less massive galaxies \mbox{($M_{\star} < 10^{10.5}M_{\odot}$)} in clusters and substructures calculated with equation \ref{quenching_efficiency} \citep{peng10} as a function of projected distance from the cluster centre or substructure centre normalised by $r_{200}$ of the cluster or substructure, accordingly. The solid orange and gold dashed lines represent the $\epsilon_{q}$ in the clusters for massive and less massive galaxies, respectively. The solid orange and dashed gold lines represent the $\epsilon_{q}$ in the substructures for massive and less massive galaxies, respectively. Left panel: spectroscopic sample. Right panel: spectrophotometric sample. The environmental quenching efficiency is higher in massive galaxies.}}
\label{fig14}
\end{figure*}

\begin{table*}
\centering
\begin{minipage}[c]{\textwidth}
\scriptsize
\caption{Mean environmental quenching efficiency in clusters and substructures.}
\begin{tabular}{lcccccc}
		\hline
		\hline
 		Environment               & \multicolumn{3}{c||}{Clusters}            		    & \multicolumn{3}{c}{Substructures}     		       \\
                                          & \multicolumn{3}{c||}{$\epsilon_{q} (\%)$}                 & \multicolumn{3}{c}{$\epsilon_{q} (\%)$}                  \\

					  & All Galaxies & Massive Galaxies & Less Massive Galaxies & All Galaxies & Massive Galaxies & Less Massive Galaxies  \\ 
		\hline                                                                                                                                         
		$r_{cl}/r_{200_{cl}}$     &              &                  &                       &              &                  &                        \\      
		Spectroscopic Sample      & 31 $\pm$ 22  & 47 $\pm$ 32      & 17 $\pm$ 5            & 25 $\pm$ 18  & 28 $\pm$ 18      & 18 $\pm$ 11            \\
		Spectrophotometric Sample & 26 $\pm$ 5   & 39 $\pm$ 26      & 24 $\pm$ 12           & 28 $\pm$ 17  & 59 $\pm$ 25      & 28 $\pm$ 18            \\
		\hline
		$r/r_{200}$               &              &		    &                       &              &                  &                        \\
		Spectroscopic Sample      & 70 $\pm$ 7   & 73 $\pm$ 11      & 55 $\pm$ 7            & 35 $\pm$ 11  & 55 $\pm$ 19      & 21 $\pm$ 4             \\
		Spectrophotometric Sample & 73 $\pm$ 5   & 92 $\pm$ 5       & 64 $\pm$ 29           & 40 $\pm$  5  & 54 $\pm$ 8       & 38 $\pm$ 4             \\
		\hline
		\hline
\vspace{-0.8cm}
\label{eq_table}
\end{tabular}
\end{minipage}
\end{table*}

\section{DISCUSSION}\label{discussion}
\subsection{Substructure detections}

The first difficulty to overcome in studies of preprocessing is the accurate detection of substructures and infalling groups. This is due mainly to the limitations given by the amount of data available, the dynamical state of the clusters and the methods used in the search and characterisation of the substructures (e.g \citealt{cohn12}, \citealt{tempel17}). 

In order to find the substructures in MACS0416 and MACS1206 we used the DS-test \citet{dressler88} along with clustering algorithms (K-means and DBSCAN; \citealt{lloyd}, \citealt{ester}). As shown in \mbox{\S \ref{ds_substructures}} we found more substructures in MACS0416 than in MASC1206. This result could be a consequence of the dynamical state of the clusters. \citet{cohn12} shows that the presence of galaxy substructures is correlated with the dynamical state of a cluster. While MACS0416 is a merger between two large structures \citep{balestra16}, MACS1206 is a relaxed cluster \citep{biviano}. However, according to \citet{tempel17} the ability to detect substructures depends on the data available regardless of the method used. In our study, as shown in \S \ref{spectroscopic_catalogues}, we have more spectroscopic data for MACS0416 than MACS1206 and this could be the reason why we have more spectroscopic members in MACS0416 than in  MACS1206 (see \ref{spec_members}), and thus more substructures in MACS0416 than MACS1206. This idea is supported by a simple test that we have done in which we have randomly selected the same number of spectroscopic members in MACS0416 and in MACS1206. In this new sample for MACS0416 we only find 10 substructures. Futhermore, the fall of spectroscopic completeness spectroscopic completeness at \mbox{$r \geq 2 \times r_{200}$} (shown in \S \ref{completeness_section}) may affect our ability of detecting substructures (\citealt{biviano}, \citealt{balestra16}). 

We note that the \mbox{$\sim$ 40\%} of our identified substructures are in the inner \mbox{($r \leq r_{200}$)} regions of the clusters, whereas the \mbox{$\sim$ 60\%} are in the outer \mbox{($r > r_{200}$)} regions (see \S \ref{ds_substructures}). Our results are in agreement with observational and theoretical studies (e.g \citealt{jaffe16}, \citealt{vijayaraghavan}, respectively) in which it is shown that most of the identifiable substructures are located in the outermost regions of galaxy clusters \mbox{(at $r > r_{200}$)}. In addition, \citet{vijayaraghavan} suggest that when a group falls into a cluster and is at a distance of \mbox{$\sim$ 2 $\times$ $r_{200}$} from the cluster centre, the group is stretched out along the direction of infall, due to the effect of the cluster gravitational potential on the group. In our case, using visual inspection (see \mbox{\S \ref{ds_substructures}}), we found that substructures in the outer regions of the clusters, especially in MACS0416, even at 2 $\times$ $r_{200}$, tend to be less elongated than substructures in the inner regions of the clusters. Interestingly, we notice that the inner \mbox{($r \leq r_{200}$)} elongated substructures found in MACS0416 are elongated along the NE-SW direction, the same in which this cluster is elongated \citep{balestra16}. This supports the notion that the inner substructures are more susceptible to the gravitational effects of the overall cluster potential.

\subsection{Colour fractions in substructures and galaxy clusters}

Figure \ref{cmd} shows that galaxies in substructures show the same bimodality in colours observed in the entire clusters. We note the absence of very bright red and blue galaxies in substructures. This could be a consequence of the selection criteria for the spectroscopic targets since the spectroscopic target selection was mostly done in a colour-colour space.

The well established morphology-density relation (\citealt{dressler80}, \citealt{postman05}) can be translated to a morphology-clustercentric radius relation (e.g \citealt{whitmore93}, \citealt{goto03}). That is, early-type galaxies are more frequent in the cluster core while late-type galaxies are more frequent in the outer regions of clusters. Moreover, the morphology of galaxies is related with other physical properties such as their stellar masses, colours and star-formation rates. In this sense, the morphology of galaxies can also be inferred from their colours, for example: early-type galaxies tend to be red, whereas late-type galaxies tend to be blue. In this context, the relation between the morphology of galaxies and the environment can be explained by a colour-environment relation (\citealt{blanton05}, \citealt{skibba09}, \citealt{grutzbauch10a}).

In the case of MACS0416 and MACS1206, the spectroscopic colour fractions as a function of projected distance from the cluster centre obtained in \S \ref{spectroscopic_fractions} (see left panel of Figure \ref{fig5}) show that galaxies in clusters and substructures follow a colour-clustercentric radius relation. Galaxies within $r_{200_{cl}}$, independent of whether they are part of a cluster or substructure, tend to be redder than galaxies beyond $r_{200_{cl}}$. This suggests that this result is likely a consequence of the effect of the cluster environment over the properties of galaxies traced by their colours (e.g \citealt{miller86}, \citealt{byrd90}, \citealt{boselli14}). Moreover, studying the colour fraction of galaxies as a function of the distance from the cluster or substructure centre (see right panel of Figure \ref{fig5}), we note that the total colour-clustercentric radius relation in substructures follows similar trends observed in the main clusters. According to this, substructures may be considered as down-scaled versions of clusters in which galaxy properties, traced by their colours, have been sculpted by the gravitational potential and physical processes within the substructures (e.g. ram-pressure stripping and galaxy-galaxy interactions), which can stop star-formation and turn galaxies red. 

Figure \ref{fig6} shows the spectroscopic colour fractions as a function of projected distance by separating galaxies in all environments according to their stellar masses in massive ($M_{\star} \geq 10^{10.5}M_{\odot}$) and less massive ($M_{\star} < 10^{10.5}M_{\odot}$). We note that fraction of green galaxies is constant in all environments, independent of the stellar mass of the galaxies and that massive galaxies tend to be red independent of the environment in which they are located. Besides, the fraction of red galaxies ($f_{r}$) in clusters and substructures at \mbox{$r < 1.5 \times r_{200_{cl}}$} is higher than the $f_{r}$ in the field, independent of the stellar mass. This result is in agreement with the result presented in the left panel of Figure 5, in which the cluster environment models galaxy properties traced by their colours at least within \mbox{$1.5 \times r_{200_{cl}}$}. Beyond \mbox{$1.5 \times r_{200_{cl}}$} $f_{r}$ in all environments become comparable. The $f_{r}$ of massive galaxies in substructures is slightly higher than in the clusters. Moreover, right panel of Figure \ref{fig6} shows that the $f_{r}$ in dense environments tend to be higher than the fraction of red galaxies in the field. This suggest that denser environments are more effective in producing red galaxies independent of the stellar masses of the galaxies. Figure \ref{fig9} shows that the results for the spectroscopic sample are reproduced by the spectrophotometric sample.

To test the colour fractions estimated using the spectroscopic sample and motivated by the low spectroscopic completeness beyond \mbox{$r > 1.5 \times r_{200}$}, we increased the statistics in the sample by adding to the spectroscopic sample z-phot selected galaxies at all radii from the cluster centres (see \S \ref{photometric_members}). For substructures, following \citet{george11}, we added $z_{phot}$ selected galaxies only within $r_{200}$ from the substructure centres (see \S \ref{photometric_substructure_members}). Figure \ref{fig8} (left panel) shows that the colour fractions as a function of the projected distance from the cluster centre obtained with the spectrophotometric sample reproduce those of the spectroscopic sample. Figure \ref{fig8} (right panel) shows that the results for the spectroscopic sample are reproduced by the spectrophotometric sample.

In the spectroscopic and background-corrected colour fractions as a function of distance (see Figures \ref{fig5} and \ref{fig8}) we see that the fraction of blue galaxies is larger in substructures than in the cluster, at all distances. Substructures host, on average, less dense environments than the entire cluster, thus allowing for a higher star formation activity (e.g. \citealt{dressler80}, \citealt{vanderwel07}). Interestingly, as noted in \citealt{perez11}, \citealt{jaffe16}, galaxy-galaxy interactions, especially those involving gas-rich galaxies, can induce bursts of star formation. This may also contribute to the enhanced blue fraction in substructures. The fraction of red galaxies in clusters and substructures is higher for massive galaxies than for less massive galaxies. This suggests that at such stellar masses internal processes play a significant role in quenching star formation.

The uncertainties in the colour fractions, in the spectroscopic sample, as a function of projected distance from the centre of the overdensity (either cluster or substructure) are of the order of \mbox{$\sim 10\%$}. We note here that our cluster sample does not include substructure members; the results of this paper, however, do not change if we consider substructure members in the cluster sample. 

With only 3 photometric bands available for MACS0416, it is not possible to obtain reliable stellar masses through SED fitting. For this reason we decided to estimate the stellar mass of galaxies using the empirical approach proposed by \citet{bell03} (see \S \ref{sec_stellar_mass}). Since the two clusters are almost at the same redshift, we used the R-band apparent magnitude as a proxy for galaxy luminosity. The colour fractions as a function of R-band luminosity and stellar mass are shown in Figures \ref{fig7} and \ref{fig10} for the spectroscopic and ``spectrophotometric'' samples, respectively. We see that the colour fractions follow the same trends with luminosity or stellar masses regardless of the galaxies being in the cluster or the substructures. 

These results support the notion that internal physical mechanisms, related to galaxy stellar mass, act independently from the environment in quenching star formation in galaxies. Interestingly, when considering the spectrophotometric sample, the red fraction is higher in substructures than in the clusters. Although still consistent within the errors, this result seems to be at odd with what we see in the spectroscopic sample, where the red fraction in substructures is slightly lower than in the clusters. We argue that this could be a consequence of the small statistics in the spectroscopic sample as we go towards the cluster outskirts, which results in the loss of substructure members. As a result, the substructure red fraction in the spectroscopic sample may be slightly underestimated. If the effect that we see in the spectrophotometric sample is real, then this suggests that the environment of substructures favours the quenching of star formation and thus pre-processing is significant in the overall quenching of star formation in clusters of galaxies. 

\subsection{Evidence for pre-processing from the colour fractions in substructures} 

An easy way to detect the existence of pre-processing in galaxy clusters consists in comparing the fractions of quiescent or star-forming galaxies in cluster outskirts and in the field. For example, \citet{wetzel13} find that galaxies in groups or satellite groups have lower SFR than in the field. \citet{hou} find that the fraction of quiescent galaxies in the outer (2 -- 3$ \times r_{200}$) substructures of galaxy clusters is higher than in the field. Similarly, \citet{haines15} show that the fraction of star-forming galaxies in clusters is lower than in the field out to 3$ \times r_{200}$ from the cluster centre. All these results point towards the existence of pre-processing.

In this work we compare the fractions of blue and red galaxies in the clusters with the fractions of star-forming and quiescent galaxies in the COSMOS/UltraVISTA field. As discussed in Section \ref{photometric_members}, we selected quiescent ans star-forming galaxies in the field with the $UVJ$ diagram. This selection cannot be done in the two clusters analysed in this work as the available photometry does not cover the rest-frame $J$ band at the redshifts of the clusters. We therefore compare the fractions of blue and red galaxies in clusters and substructures with the fractions of quiescent and star-forming galaxies in the field. This comparison can be done as the fractions of blue and red galaxies in clusters and substructures are corrected for contamination of dusty star-forming galaxies as discussed in Section \S \ref{photometric_sample}.

Figure \ref{fig5} shows that in the spectroscopic sample the fractions of red galaxies remain higher than the field in the cluster and substructures out to \mbox{2$ \times r_{200_{cl}}$} from the cluster centres and 1$ \times r_{200}$ from the overdensity centre. At larger distances from the cluster centre (\mbox{$r > 2 \times r_{200_{cl}}$}) the fraction of quiescent galaxies in the field becomes slightly higher than in substructures. We remind here that the spectroscopic completeness of the sample is a decreasing function of cluster-centric radius (Figure \ref{fig1}), implying that results at projected distances greater than 2$ \times r_{200}$ may be affected by sample incompleteness even after weighting the observed colour fractions. However, considering the uncertainties on colour fractions, the fraction of red galaxies in substructures is at least comparable to that in the field out to \mbox{2.5$ \times r_{200_{cl}}$} from the cluster centre. Figure \ref{fig8} shows, indeed, that when considering also $z_{phot}$-selected members the fraction of red galaxies in substructures is higher or at least comparable to that in the field out to \mbox{2.5$ \times r_{200_{cl}}$} from the cluster centre. In the case of colour fractions as a function of projected distance from the centre of an overdensity, the right panel of Figures \ref{fig5} and \ref{fig8} show that the fraction of red galaxies in substructures is higher than that in the field. 

We conclude from this that the trends of the colour fractions with projected distance for cluster and substructure galaxies agree with the existence of pre-processing in galaxy clusters.

\citet{bianconi17} studied the star formation in infalling substructures for a sample of clusters at \mbox{0.15 $<~z~<$ 0.3} selected from the Local Cluster Substructure Survey (LoCuSS, \citealt{smith10}). They find that the fraction of star-forming galaxies in the substructures is always lower than in the clusters, interpreting this as evidence of pre-processing. Although this result appears in disagreement with our Figures \ref{fig5} and \ref{fig8}, we can reconcile our findings with those of \citet{bianconi17} by noting that those authors identified substructures using the X-ray luminosity distributions of the clusters. Such a method privileges the selection of the most massive and dense substructures, where the environmental quenching effects are stronger. We also point out that the clusters in \citet{bianconi17} are at lower redshifts than ours and thus their substructures are likely to host larger fractions of passive galaxies (see e.g. \citealt{li09}). Finally, these authors restricted themselves only to massive galaxies (\mbox{$M_{\star} \geq 2 \times 10^{10} M_{\odot}$}), which may be more affected by internal quenching processes (see e.g. \citealt{muzzin}, \citealt{gavazzi15}, also Figures \ref{fig7} and \ref{fig10} in this paper). Our results (see Figures \ref{fig6} and \ref{fig9}) show that, considering the uncertainties, the fraction of red massive \mbox{($M_{\star} \geq 10^{10.5}M_{\odot}$)} galaxies in substructures is higher or at least comparable to the fraction of red massive galaxies in the clusters as a function of projected distance from the cluster centre. This result is in agreement with \citet{bianconi17}. These results are in agreement with some authors (e.g \citealt{grutzbauch10b}) that find that the effects of stellar mass and environments in producing red galaxies are not independent and that massive galaxies are preferentially located in denser environments. 

Figure \ref{fig11} shows that the environmental quenching efficiency in substructures is lower than in the main cluster and begins to be comparable to the main cluster at \mbox{$r >$ 1$ \times r_{200_{cl}}$}. Figure \ref{fig13} shows that the environmental quenching efficiency in clusters is always higher than in the substructures. In particular, we note that while $\epsilon_{q}$ increases towards the cluster centre, it remains approximately constant with the distance from the centre of the substructures. Despite these differences in the spatial distribution of the quenching efficiency, we find that on average $\epsilon_{q}$ has similar values in the main cluster and in the substructures (see Table \ref{eq_table}). All these results can be explained by the fact that environmental quenching processes are stronger in cluster cores than in substructures. Interestingly, we note that the $\epsilon_{q}$ of massive galaxies ($M_{\star} \geq 10^{10.5}M_{\odot}$) in substructures is higher than $\epsilon_{q}$ of the massive galaxies in the cluster (see right panel of Figure \ref{fig12}). Our results could be explain by the fact that in lower-mass environments such as galaxy groups or substructures is more likely to produce red massive galaxies by galaxy mergers than in clusters \citet{mihos03}. This fact could be indicating that internal and external mechanisms act together in producing red galaxies in substructures. 

We note in Figure \ref{fig14} that the values of $\epsilon_{q}$ for massive and less massive galaxies in the inner regions ($ < r_{200}$) of clusters and substructures are consistent between them. However, $\epsilon_{q}$ appears larger for massive galaxies in both the cluster and substructure samples. We need a larger sample to test such a result, which will be addressed in a forthcoming paper.

\citet{kawinwanichakij17}, using galaxies at \mbox{$0.5 < z < 2.0$} from the FourStar Galaxy Evolution Survey (ZFOURGE \citealt{straatman16}), found that in the mass range of \mbox{8.8 $< \log(M/M_{\odot}) <$ 10.0} the environmental quenching efficiency is \mbox{$\sim$ 30\%} at \mbox{$z < 1$}. These authors suggest that, at \mbox{$z < 1$}, environmental quenching plays a key role in suppressing the star-formation in low mass galaxies. In our case, using only the spectroscopic sample, we estimate a mean quenching efficiency of \mbox{$\epsilon_{q}$ = (31 $\pm$ 22)\%} and \mbox{$\epsilon_{q}$ = (25 $\pm$ 18)\%} for clusters and substructures, respectively (see Table \ref{eq_table}). When we use the spectrophotometric sample we obtain a mean quenching efficiency of \mbox{$\epsilon_{q}$ = (26 $\pm$ 16)\%} for clusters, and a \mbox{$\epsilon_{q}$ = (28 $\pm$ 17)\%} for substructures (see Table \ref{eq_table}). Our values are close to those obtained by \citet{kawinwanichakij17}. Besides, we note that for less massive galaxies \mbox{($M_{\star} < 10^{10.5}M_{\odot}$)} in substructures we found a mean environmental quenching efficiency of the order of \mbox{(18 $\pm$ 11)\%} and \mbox{(25 $\pm$ 18)\%} for the spectroscopic and spectrophotometric samples, respectively. Our mean values are slightly lower than the value found by \citet{kawinwanichakij17}. However, in the case of massive galaxies \mbox{($M_{\star} \geq 10^{10.5}M_{\odot}$)} we found a mean environmental quenching efficiency in substructures of \mbox{$\epsilon_{q}$ = (59 $\pm$ 25)\%} for the spectrophotometric sample, this value is higher than the environmental quenching efficiency of massive galaxies in the cluster (see Table \ref{eq_table}). This suggest that in our studied clusters the environmental quenching plays a role in producing red galaxies in all stellar mass ranges. 

\section{SUMMARY AND CONCLUSIONS}\label{conslusion}

We have studied the effect of the local environment on the colour of galaxies in two clusters at \mbox{$z \sim 0.4$} drawn from the CLASH-VLT survey \citep{rosati}. In these clusters, we have identified several substructures using a combination of statistical methods (DS-test) and clustering algorithms (DBSCAN). Most of the identified substructures are located in the cluster outskirts (\mbox{$r \geq r_{200}$}), in agreement with literature results.  

We have investigated the spectroscopic and background-corrected colour fraction as a function of projected distance from the cluster centre for galaxies in cluster and substructures, and as a function of projected distance from the centre of the overdensity (either cluster or substructure) for clusters and substructures, accordingly. We found that the colour-clustercentric radius relation is well established both in clusters and substructures. In addition we found that the colour-clustercentric radius relation in the clusters can be reproduced by the colour-clustercentric radius relation in substructures. This suggests that substructures may be considered as down-scaled versions of clusters, in which the internal physical properties of the substructures sculpt galaxy properties such as colours, SFR and morphologies. 

We find that the fraction of blue galaxies in both clusters and substructures is lower than the fraction of star-forming galaxies in the field. The fractions of red and blue galaxies in substructures are intermediate between that of the main cluster and the fractions of quiescent and star-forming galaxies in the field. This result supports the notion of the existence of pre-processing in galaxy clusters. 

The environmental quenching efficiency in the centres of substructures is lower than in the centres of clusters and becomes comparable to that of the clusters when the distance from the centre of the overdensity increases. We also find that the environmental quenching efficiency of external substructures (\mbox{$r \geq r_{200}$} from the cluster centres) is comparable to that of the main cluster. The average environmental quenching efficiencies of cluster and substructures are similar. 

We find that massive galaxies \mbox{($M_{\star} \geq 10^{10.5}M_{\odot}$)} tend to be redder than less massive galaxies \mbox{($M_{\star} < 10^{10.5}M_{\odot}$)} in all environments. However the environmental quenching efficiency of massive galaxies in substructures is higher than the environmental quenching efficiency of galaxies in clusters. This fact suggests that stellar mass and environment play a combined role in producing red galaxies in substructures.

The analysis of MACS0416 and MACS1206 shows that the study of substructures in clusters is fundamental to understand the evolution of galaxies in dense environments. Clusters assembled a significant part of their mass through the accretion of smaller groups, and the properties of galaxies in clusters bear the imprint of this assembly history as we show in this paper. The analysis developed in this work will be extended to the entire CLASH-VLT sample; this work is already ongoing and will allow us to study galaxy pre-processing with high statistical significance in a large sample of clusters.

\section*{ACKNOWLEDGEMENTS}
We thank the anonymous referee for the useful comments that greatly improved this paper. We thank D. Gruen for his constructive feedback and suggestions about the use of MACS0416's catalogue. D.O-R acknowledges the financial support provided by CONICYT-PCHA through a PhD Scholarship, `Beca Doctorado Nacional A\~{n}o 2015', under contract 2015-21150415. P.C. acknowledges the support provided by FONDECYT postdoctoral research grant no 3160375. R.D. gratefully acknowledges the support provided by the BASAL Center for Astrophysics and Associated Technologies (CATA). Y. J. acknowledges support from CONICYT PAI (Concurso Nacional de Inserci\'{o}n en la Academia 2017) No. 79170132. This work was made using data taken under the ESO programmes ID 094.A-0115(B) and ID 094.A-0525(A), and also 186.A-0798. This study is based on a K$_{s}$-selected catalog of the COSMOS/UltraVISTA survey.

\bibliographystyle{mn2e}
\bibliography{biblo.bib}

\appendix

\section{Data Tables}

Here we present the data use to plot the colour fractions and quenching efficiency in spectroscopic and spectrophotometric samples. Tables \ref{tablea1} -- \ref{tablea6} show the spectroscopic colour fractions as a function of projected distance, magnitude and stellar masses for substructures and clusters. Tables \ref{tablea7} -- \ref{tablea12}. Tables \ref{tablea13} -- \ref{tablea16} show the mean environmental quenching efficiency as a function of distance for spectroscopic and spectrophotometric samples.

\begin{table*}
\centering
\begin{threeparttable}
	\caption{Spectroscopic colour fraction as a function of distance from the cluster centre.}
  	\label{tablea1}
	{\small
 		\begin{tabular}{lccccc}
		\hline
		\hline
 			& \multicolumn{4}{c}{$r_{cl}/r_{200_{cl}}$}         		       \\
				& 0.5                       & 1.5                        & 2.5                       & 3.5                       \\
		\hline
		Cluster         &                           &                            &                           &                           \\
		$f_b$		& 0.209$_{-0.007}^{+0.007}$ & 0.588$_{-0.013}^{+0.013}$  & 0.706$_{-0.026}^{+0.024}$ & 0.321$_{-0.092}^{+0.194}$ \\
		$f_g$		& 0.139$_{-0.006}^{+0.006}$ & 0.154$_{-0.009}^{+0.010}$  & 0.158$_{-0.018}^{+0.022}$ & 0.250$_{-0.092}^{+0.194}$ \\
		$f_r$		& 0.651$_{-0.009}^{+0.008}$ & 0.249$_{-0.011}^{+0.012}$  & 0.131$_{-0.018}^{+0.022}$ & 0.415$_{-0.158}^{+0.158}$ \\
		\hline
		\hline
		Substructures  &			    &                            &                           &                           \\
		$f_b$		& 0.310$_{-0.024}^{+0.029}$ & 0.614$_{-0.024}^{+0.023}$  & 0.905$_{-0.074}^{+0.033}$ & -                         \\
		$f_g$		& 0.177$_{-0.021}^{+0.027}$ & 0.178$_{-0.017}^{+0.020}$  & 0.033$_{-0.019}^{+0.014}$ & -                         \\
		$f_r$		& 0.495$_{-0.032}^{+0.030}$ & 0.202$_{-0.020}^{+0.022}$  & 0.092$_{-0.033}^{+0.074}$ & -                         \\
		\hline
		\hline
		\end{tabular}  
}
\end{threeparttable}
\end{table*}

\begin{table*}
\centering
\begin{threeparttable}
	\caption{Spectroscopic colour fraction as a function of distance from the overdensity centre}
  	\label{tablea2}
	{\small
 		\begin{tabular}{lcccccc}
		\hline
		\hline
 			        & \multicolumn{5}{c}{$r/r_{200}$}         		                                                                                     \\
				& 0.1                       & 0.3                        & 0.5                       & 0.7                       & 0.9                       \\
		\hline
		Cluster         &                           &                            &                           &                           &                           \\
		$f_b$		& 0.104$_{-0.008}^{+0.011}$ & 0.153$_{-0.011}^{+0.014}$  & 0.197$_{-0.019}^{+0.023}$ & 0.237$_{-0.022}^{+0.027}$ & 0.380$_{-0.027}^{+0.029}$ \\
		$f_g$		& 0.124$_{-0.012}^{+0.014}$ & 0.110$_{-0.011}^{+0.013}$  & 0.137$_{-0.017}^{+0.022}$ & 0.194$_{-0.022}^{+0.026}$ & 0.137$_{-0.018}^{+0.023}$ \\
		$f_r$		& 0.770$_{-0.016}^{+0.014}$ & 0.734$_{-0.017}^{+0.016}$  & 0.664$_{-0.028}^{+0.025}$ & 0.567$_{-0.031}^{+0.029}$ & 0.482$_{-0.029}^{+0.029}$ \\
		\hline
		\hline 
		Substructures   &			    &                            &                           &                           &                           \\
		$f_b$		& 0.375$_{-0.024}^{+0.051}$ & 0.423$_{-0.055}^{+0.111}$  & 0.410$_{-0.039}^{+0.183}$ & 0.462$_{-0.031}^{+0.159}$ & 0.734$_{-0.103}^{+0.274}$ \\
		$f_g$		& 0.200$_{-0.061}^{+0.116}$ & 0.074$_{-0.055}^{+0.111}$  & 0.067$_{-0.050}^{+0.101}$ & 0.056$_{-0.042}^{+0.086}$ & 0.129$_{-0.096}^{+0.179}$ \\
		$f_r$		& 0.349$_{-0.116}^{+0.061}$ & 0.385$_{-0.111}^{+0.055}$  & 0.450$_{-0.183}^{+0.039}$ & 0.402$_{-0.159}^{+0.031}$ & 0.119$_{-0.274}^{+0.103}$ \\
		\hline
		\hline
		\end{tabular}  
}
\end{threeparttable}
\end{table*}

\begin{table*}
\centering
\begin{threeparttable}
	\caption{Spectroscopic colour fraction as a function of distance from the cluster centre for massive and less massive galaxies}
  	\label{tablea3}
	{\small
 		\begin{tabular}{lccccc}
		\hline
		\hline
 			& \multicolumn{4}{c}{$r_{cl}/r_{200_{cl}}$}         		       \\
				& 0.5                       & 1.5                        & 2.5                       & 3.5                        \\
		\hline
		Cluster         &                           &                            &                           &                            \\
                                & \multicolumn{4}{c}{$M_{\star} \geq 10^{10.5}M_{\odot}$}                                                         \\
		                &			    &                            &                           &                           \\

		$f_b$		& 0.071$_{-0.006}^{+0.007}$ & 0.239$_{-0.023}^{+0.032}$  & 0.243$_{-0.028}^{+0.037}$ & 0.252$_{-0.096}^{+0.179}$  \\
		$f_g$		& 0.050$_{-0.008}^{+0.009}$ & 0.105$_{-0.020}^{+0.030}$  & 0.353$_{-0.039}^{+0.043}$ & 0.129$_{-0.100}^{+0.179}$  \\
		$f_r$		& 0.878$_{-0.012}^{+0.009}$ & 0.633$_{-0.038}^{+0.031}$  & 0.391$_{-0.042}^{+0.043}$ & 0.723$_{-0.179}^{+0.096}$  \\
		                &			    &                            &                           &                           \\
                                & \multicolumn{4}{c}{$M_{\star} < 10^{10.5}M_{\odot}$}                                                            \\
		                &			    &                            &                           &                           \\
		$f_b$		& 0.301$_{-0.010}^{+0.011}$ & 0.678$_{-0.014}^{+0.014}$   & 0.822$_{-0.026}^{+0.021}$ & 0.518$_{-0.203}^{+0.203}$ \\
		$f_g$		& 0.197$_{-0.009}^{+0.100}$ & 0.167$_{-0.010}^{+0.011}$   & 0.109$_{-0.016}^{+0.022}$ & 0.500$_{-0.203}^{+0.129}$ \\
		$f_r$		& 0.501$_{-0.012}^{+0.012}$ & 0.150$_{-0.011}^{+0.012}$   & 0.067$_{-0.014}^{+0.020}$ & 0.107$_{-0.096}^{+0.179}$ \\

		\hline
		\hline
		Substructures   &			    &                            &                           &                           \\
                                & \multicolumn{4}{c}{$M_{\star} \geq 10^{10.5}M_{\odot}$}                                                        \\
		                &			    &                            &                           &                           \\
		$f_b$		& 0.130$_{-0.006}^{+0.012}$ & 0.256$_{-0.048}^{+0.113}$  & 0.393$_{-0.210}^{+0.024}$ & -                         \\
		$f_g$		& 0.130$_{-0.027}^{+0.043}$ & 0.100$_{-0.034}^{+0.107}$  & 0.293$_{-0.210}^{+0.309}$ & -                         \\
		$f_r$		& 0.722$_{-0.043}^{+0.027}$ & 0.623$_{-0.117}^{+0.071}$  & 0.587$_{-0.309}^{+0.210}$ & -                         \\
		                &			    &                            &                           &                           \\
                                & \multicolumn{4}{c}{$M_{\star} < 10^{10.5}M_{\odot}$}                                                           \\
		                &			    &                            &                           &                           \\
		$f_b$		& 0.420$_{-0.036}^{+0.040}$ & 0.696$_{-0.026}^{+0.240}$  & 0.982$_{-0.034}^{+0.016}$ & -                         \\
		$f_g$		& 0.205$_{-0.029}^{+0.036}$ & 0.195$_{-0.020}^{+0.023}$  & 0.021$_{-0.016}^{+0.034}$ & -                         \\
		$f_r$		& 0.362$_{-0.039}^{+0.040}$ & 0.105$_{-0.016}^{+0.020}$  & 0.017$_{-0.016}^{+0.034}$ & -                         \\
		\hline
		\hline
		\end{tabular}  
}
\end{threeparttable}
\end{table*}

\begin{table*}
\centering
\begin{threeparttable}
	\caption{Spectroscopic colour fraction as a function of distance from the overdensity centre for massive and less massive galaxies}
  	\label{tablea4}
	{\small
 		\begin{tabular}{lcccccc}
		\hline
		\hline
 			        & \multicolumn{5}{c}{$r/r_{200}$}         		                                                                                     \\
				& 0.1                       & 0.3                        & 0.5                       & 0.7                       & 0.9                       \\
		\hline
		Cluster         &                           &                            &                           &                           &                           \\
                                & \multicolumn{5}{c}{$M_{\star} \geq 10^{10.5}M_{\odot}$}                                                                                    \\
		                &                           &                            &                           &                           &                           \\
		$f_b$		& 0.051$_{-0.004}^{+0.008}$ & 0.062$_{-0.008}^{+0.012}$  & 0.089$_{-0.012}^{+0.014}$ & 0.052$_{-0.006}^{+0.012}$ & 0.147$_{-0.026}^{+0.047}$ \\
		$f_g$		& 0.032$_{-0.012}^{+0.017}$ & 0.041$_{-0.014}^{+0.018}$  & 0.059$_{-0.011}^{+0.018}$ & 0.065$_{-0.017}^{+0.036}$ & 0.081$_{-0.022}^{+0.044}$ \\
		$f_r$		& 0.920$_{-0.017}^{+0.012}$ & 0.899$_{-0.020}^{+0.016}$  & 0.852$_{-0.021}^{+0.016}$ & 0.888$_{-0.036}^{+0.017}$ & 0.770$_{-0.054}^{+0.037}$ \\
		                &                           &                            &                           &                           &                           \\
                                & \multicolumn{5}{c}{$M_{\star} < 10^{10.5}M_{\odot}$}                                                                                    \\
		                &                           &                            &                           &                           &                           \\
		$f_b$		& 0.147$_{-0.014}^{+0.018}$ & 0.245$_{-0.021}^{+0.024}$  & 0.285$_{-0.021}^{+0.024}$ & 0.338$_{-0.031}^{+0.036}$ & 0.460$_{-0.033}^{+0.037}$ \\
		$f_g$		& 0.198$_{-0.019}^{+0.022}$ & 0.181$_{-0.019}^{+0.023}$  & 0.200$_{-0.027}^{+0.034}$ & 0.261$_{-0.030}^{+0.036}$ & 0.156$_{-0.016}^{+0.019}$ \\
		$f_r$		& 0.653$_{-0.025}^{+0.026}$ & 0.572$_{-0.027}^{+0.026}$  & 0.514$_{-0.039}^{+0.038}$ & 0.399$_{-0.036}^{+0.038}$ & 0.383$_{-0.023}^{+0.024}$ \\
		\hline
		\hline 
		Substructures   &			    &                            &                           &                           &                           \\
                                & \multicolumn{5}{c}{$M_{\star} \geq 10^{10.5}M_{\odot}$}                                                                                    \\
		                &                           &                            &                           &                           &                           \\
		$f_b$		& 0.145$_{-0.024}^{+0.052}$ & 0.205$_{-0.056}^{+0.111}$  & 0.237$_{-0.039}^{+0.183}$ & 0.219$_{-0.031}^{+0.159}$ & 0.356$_{-0.103}^{+0.274}$ \\
		$f_g$		& 0.200$_{-0.061}^{+0.116}$ & 0.074$_{-0.055}^{+0.111}$  & 0.067$_{-0.050}^{+0.101}$ & 0.056$_{-0.042}^{+0.086}$ & 0.129$_{-0.096}^{+0.179}$ \\
		$f_r$		& 0.664$_{-0.116}^{+0.061}$ & 0.768$_{-0.111}^{+0.055}$  & 0.738$_{-0.183}^{+0.039}$ & 0.755$_{-0.159}^{+0.031}$ & 0.623$_{-0.274}^{+0.103}$ \\
		                &                           &                            &                           &                           &                           \\
                                & \multicolumn{5}{c}{$M_{\star} < 10^{10.5}M_{\odot}$}                                                                                    \\
		                &                           &                            &                           &                           &                           \\
		$f_b$		& 0.469$_{-0.063}^{+0.068}$ & 0.535$_{-0.056}^{+0.055}$  & 0.514$_{-0.062}^{+0.064}$ & 0.584$_{-0.054}^{+0.052}$ & 0.825$_{-0.055}^{+0.037}$ \\
		$f_g$		& 0.286$_{-0.052}^{+0.067}$ & 0.250$_{-0.042}^{+0.054}$  & 0.200$_{-0.042}^{+0.061}$ & 0.182$_{-0.034}^{+0.048}$ & 0.176$_{-0.037}^{+0.055}$ \\
		$f_r$		& 0.237$_{-0.052}^{+0.067}$ & 0.208$_{-0.042}^{+0.054}$  & 0.277$_{-0.055}^{+0.065}$ & 0.226$_{-0.042}^{+0.052}$ & 0.008$_{-0.007}^{+0.016}$ \\

		\hline
		\hline
		\end{tabular}  
}
\end{threeparttable}
\end{table*}

\begin{table*}
\centering
\begin{threeparttable}
	\caption{Spectroscopic colour fraction as a function of $R_{c}$}
  	\label{tablea5}
	{\small
 		\begin{tabular}{lcccccc}
		\hline
		\hline
 			        & \multicolumn{5}{c}{$R_{c}$}         		                                                                                     \\
				& 19.5                      & 20.5                        & 21.5                       & 22.5                       & 23.5                       \\
		\hline
		Cluster         &                           &                            &                           &                           &                           \\
		$f_b$		& 0.189$_{-0.017}^{+0.034}$ & 0.240$_{-0.015}^{+0.018}$  & 0.382$_{-0.016}^{+0.017}$ & 0.454$_{-0.017}^{+0.018}$ & 0.578$_{-0.023}^{+0.023}$ \\
		$f_g$		& 0.048$_{-0.019}^{+0.025}$ & 0.121$_{-0.014}^{+0.018}$  & 0.121$_{-0.011}^{+0.013}$ & 0.171$_{-0.013}^{+0.014}$ & 0.192$_{-0.017}^{+0.019}$ \\
		$f_r$		& 0.744$_{-0.040}^{+0.023}$ & 0.617$_{-0.022}^{+0.020}$  & 0.481$_{-0.018}^{+0.017}$ & 0.362$_{-0.018}^{+0.018}$ & 0.222$_{-0.019}^{+0.021}$ \\
		\hline
		\hline 
		Substructures   &			    &                            &                           &                           &                           \\
		$f_b$		& 0.1805$_{-0.034}^{+0.070}$ & 0.210$_{-0.026}^{+0.045}$  & 0.422$_{-0.045}^{+0.051}$ & 0.553$_{-0.046}^{+0.046}$ & 0.770$_{-0.046}^{+0.036}$ \\
		$f_g$		& 0.045$_{-0.036}^{+0.051}$ & 0.185$_{-0.036}^{+0.051}$  & 0.167$_{-0.031}^{+0.045}$ & 0.205$_{-0.032}^{+0.042}$ & 0.167$_{-0.030}^{+0.042}$ \\
		$f_r$		& 0.955$_{-0.070}^{+0.034}$ & 0.704$_{-0.056}^{+0.045}$  & 0.479$_{-0.050}^{+0.051}$ & 0.234$_{-0.038}^{+0.045}$ & 0.061$_{-0.018}^{+0.034}$ \\
		\hline
		\hline
		\end{tabular}  
}
\end{threeparttable}
\end{table*}

\begin{table*}
\centering
\begin{threeparttable}
	\caption{Spectroscopic colour fraction as a function of $\log(M_{\star}/M_{\odot})$}
  	\label{tablea6}
	{\small
 		\begin{tabular}{lcccc}
		\hline
		\hline
		                & \multicolumn{4}{c}{$\log(M_{\star}/M_{\odot})$}		                                                \\
		\hline
				& 8.5	                    & 9.5                 	& 10.5	                    & 11.5	        	\\
		\hline
		Cluster 	&                           &                           &                           &		                \\
		$f_b$		& 0.978$_{-0.041}^{+0.019}$ & 0.636$_{-0.016}^{+0.015}$ & 0.296$_{-0.010}^{+0.011}$ & 0.141$_{-0.003}^{+0.006}$	\\
		$f_g$		& 0.025$_{-0.019}^{+0.041}$ & 0.182$_{-0.117}^{+0.013}$ & 0.139$_{-0.009}^{+0.010}$ & 0.026$_{-0.010}^{+0.014}$	\\
		$f_r$		& 0.021$_{-0.019}^{+0.041}$ & 0.175$_{-0.012}^{+0.014}$ & 0.546$_{-0.013}^{+0.012}$ & 0.808$_{-0.014}^{+0.010}$	\\
		\hline
		\hline		
		Substructures	&                           &                           &                           &		                \\
		$f_b$		& -                         & 0.764$_{-0.029}^{+0.025}$ & 0.323$_{-0.022}^{+0.027}$ & 0.156$_{-0.011}^{+0.018}$ \\
		$f_g$		& -                         & 0.241$_{-0.031}^{+0.037}$ & 0.177$_{-0.026}^{+0.034}$ & 0.582$_{-0.040}^{+0.038}$ \\
		$f_r$		& -                         & 0.048$_{-0.013}^{+0.028}$ & 0.483$_{-0.013}^{+0.028}$ & 0.816$_{-0.028}^{+0.013}$ \\
		\hline
		\hline
		\end{tabular}  
}
\end{threeparttable}
\end{table*}

\begin{table*}
\centering
\begin{threeparttable}
	\caption{Spectrophotometric colour fraction as a function of distance from the cluster centre.}
  	\label{tablea7}
	{\small
 		\begin{tabular}{lccccc}
		\hline
		\hline
 			& \multicolumn{4}{c}{$r_{cl}/r_{200_{cl}}$}         		       \\
				& 0.5                       & 1.5                        & 2.5                       & 3.5                       \\
		\hline
		Cluster         &                           &                            &                           &                           \\
		$f_b$		& 0.290$_{-0.013}^{+0.013}$ & 0.640$_{-0.013}^{+0.013}$  & 0.674$_{-0.012}^{+0.012}$ & 0.664$_{-0.020}^{+0.020}$ \\
		$f_g$		& 0.089$_{-0.008}^{+0.008}$ & 0.121$_{-0.008}^{+0.009}$  & 0.121$_{-0.008}^{+0.008}$ & 0.093$_{-0.011}^{+0.013}$ \\
		$f_r$		& 0.627$_{-0.014}^{+0.014}$ & 0.238$_{-0.012}^{+0.012}$  & 0.206$_{-0.010}^{+0.011}$ & 0.246$_{-0.018}^{+0.019}$ \\
		\hline
		\hline
		Substructures  &			    &                            &                           &                           \\
		$f_b$		& 0.416$_{-0.030}^{+0.031}$ & 0.650$_{-0.026}^{+0.025}$  & 0.853$_{-0.179}^{+0.114}$ & -                         \\
		$f_g$		& 0.099$_{-0.018}^{+0.020}$ & 0.077$_{-0.013}^{+0.014}$  & 0.248$_{-0.104}^{+0.129}$ & -                         \\
		$f_r$		& 0.481$_{-0.032}^{+0.031}$ & 0.276$_{-0.024}^{+0.024}$  & 0.089$_{-0.078}^{+0.154}$ & -                         \\
		\hline
		\hline
		\end{tabular}  
}
\end{threeparttable}
\end{table*}

\begin{table*}
\centering
\begin{threeparttable}
	\caption{Spectrophotometric colour fraction as a function of distance from the overdensity centre}
  	\label{tablea8}
	{\small
 		\begin{tabular}{lcccccc}
		\hline
		\hline
 			        & \multicolumn{5}{c}{$r/r_{200}$}         		                                                                                     \\
				& 0.1                       & 0.3                        & 0.5                       & 0.7                       & 0.9                       \\
		\hline
		Cluster         &                           &                            &                           &                           &                           \\
		$f_b$		& 0.067$_{-0.015}^{+0.022}$ & 0.212$_{-0.026}^{+0.029}$  & 0.282$_{-0.030}^{+0.032}$ & 0.342$_{-0.033}^{+0.034}$ & 0.463$_{-0.032}^{+0.032}$ \\
		$f_g$		& 0.042$_{-0.013}^{+0.017}$ & 0.074$_{-0.016}^{+0.019}$  & 0.108$_{-0.020}^{+0.023}$ & 0.097$_{-0.018}^{+0.020}$ & 0.115$_{-0.019}^{+0.020}$ \\
		$f_r$		& 0.892$_{-0.027}^{+0.023}$ & 0.751$_{-0.032}^{+0.030}$  & 0.642$_{-0.035}^{+0.034}$ & 0.592$_{-0.034}^{+0.032}$ & 0.454$_{-0.031}^{+0.031}$ \\
		\hline
		\hline 
		Substructures   &			    &                            &                           &                           &                           \\
		$f_b$		& 0.548$_{-0.128}^{+0.134}$ & 0.571$_{-0.060}^{+0.060}$  & 0.574$_{-0.044}^{+0.045}$ & 0.584$_{-0.040}^{+0.039}$ & 0.586$_{-0.038}^{+0.038}$ \\
		$f_g$		& 0.045$_{-0.030}^{+0.051}$ & 0.106$_{-0.030}^{+0.037}$  & 0.086$_{-0.022}^{+0.026}$ & 0.093$_{-0.020}^{+0.024}$ & 0.064$_{-0.016}^{+0.020}$ \\
		$f_r$		& 0.460$_{-0.098}^{+0.097}$ & 0.349$_{-0.055}^{+0.056}$  & 0.354$_{-0.043}^{+0.043}$ & 0.334$_{-0.038}^{+0.038}$ & 0.359$_{-0.037}^{+0.037}$ \\
		\hline
		\hline
		\end{tabular}  
}
\end{threeparttable}
\end{table*}

\begin{table*}
\centering
\begin{threeparttable}
	\caption{Spectrophotometric colour fraction as a function of distance from the cluster centre for massive and less massive galaxies}
  	\label{tablea9}
	{\small
 		\begin{tabular}{lccccc}
		\hline
		\hline
 			& \multicolumn{4}{c}{$r_{cl}/r_{200_{cl}}$}         		       \\
				& 0.5                       & 1.5                        & 2.5                       & 3.5                        \\
		\hline
		Cluster         &                           &                            &                           &                            \\
                                & \multicolumn{4}{c}{$M_{\star} \geq 10^{10.5}M_{\odot}$}                                                         \\
		                &			    &                            &                           &                           \\

		$f_b$		& 0.083$_{-0.009}^{+0.010}$ & 0.301$_{-0.028}^{+0.029}$  & 0.272$_{-0.026}^{+0.028}$ & 0.193$_{-0.035}^{+0.055}$  \\
		$f_g$		& 0.041$_{-0.010}^{+0.011}$ & 0.105$_{-0.020}^{+0.023}$  & 0.157$_{-0.025}^{+0.028}$ & 0.060$_{-0.033}^{+0.054}$  \\
		$f_r$		& 0.902$_{-0.015}^{+0.013}$ & 0.608$_{-0.035}^{+0.034}$  & 0.584$_{-0.037}^{+0.036}$ & 0.779$_{-0.097}^{+0.046}$  \\
		                &			    &                            &                           &                           \\
                                & \multicolumn{4}{c}{$M_{\star} < 10^{10.5}M_{\odot}$}                                                            \\
		                &			    &                            &                           &                           \\
		$f_b$		& 0.393$_{-0.018}^{+0.017}$ & 0.697$_{-0.013}^{+0.014}$  & 0.724$_{-0.012}^{+0.011}$ & 0.691$_{-0.020}^{+0.021}$ \\
		$f_g$		& 0.112$_{-0.011}^{+0.012}$ & 0.125$_{-0.010}^{+0.009}$  & 0.117$_{-0.008}^{+0.008}$ & 0.095$_{-0.011}^{+0.013}$ \\
		$f_r$		& 0.506$_{-0.018}^{+0.018}$ & 0.183$_{-0.012}^{+0.012}$  & 0.164$_{-0.009}^{+0.011}$ & 0.224$_{-0.018}^{+0.019}$ \\

		\hline
		\hline
		Substructures   &			    &                            &                           &                           \\
                                & \multicolumn{4}{c}{$M_{\star} \geq 10^{10.5}M_{\odot}$}                                                        \\
		                &			    &                            &                           &                           \\
		$f_b$		& 0.162$_{-0.017}^{+0.036}$ & 0.307$_{-0.063}^{+0.076}$  & 0.368$_{-0.210}^{+0.307}$ & -                         \\
		$f_g$		& 0.024$_{-0.017}^{+0.036}$ & 0.047$_{-0.032}^{+0.056}$  & 0.294$_{-0.210}^{+0.307}$ & -                         \\
		$f_r$		& 0.811$_{-0.038}^{+0.018}$ & 0.697$_{-0.096}^{+0.082}$  & 0.432$_{-0.348}^{+0.330}$ & -                         \\
		                &			    &                            &                           &                           \\
                                & \multicolumn{4}{c}{$M_{\star} < 10^{10.5}M_{\odot}$}                                                           \\
		                &			    &                            &                           &                           \\
		$f_b$		& 0.453$_{-0.033}^{+0.034}$ & 0.680$_{-0.020}^{+0.022}$  & 0.879$_{-0.034}^{+0.034}$ & -                         \\
		$f_g$		& 0.638$_{-0.025}^{+0.026}$ & 0.080$_{-0.013}^{+0.015}$  & 0.295$_{-0.024}^{+0.024}$ & -                         \\
		$f_r$		& 0.434$_{-0.166}^{+0.099}$ & 0.245$_{-0.111}^{+0.138}$  & 0.095$_{-0.084}^{+0.181}$ & -                         \\
		\hline
		\hline
		\end{tabular}  
}
\end{threeparttable}
\end{table*}

\begin{table*}
\centering
\begin{threeparttable}
	\caption{Spectrophotometric colour fraction as a function of distance from the overdensity centre for massive and less massive galaxies}
  	\label{tablea10}
	{\small
 		\begin{tabular}{lcccccc}
		\hline
		\hline
 			        & \multicolumn{5}{c}{$r/r_{200}$}         		                                                                                     \\
				& 0.1                       & 0.3                        & 0.5                       & 0.7                       & 0.9                       \\
		\hline
		Cluster         &                           &                            &                           &                           &                           \\
                                & \multicolumn{5}{c}{$M_{\star} \geq 10^{10.5}M_{\odot}$}                                                                                    \\
		                &                           &                            &                           &                           &                           \\
		$f_b$		& 0.056$_{-0.006}^{+0.014}$ & 0.067$_{-0.011}^{+0.017}$  & 0.096$_{-0.021}^{+0.029}$ & 0.074$_{-0.015}^{+0.024}$ & 0.184$_{-0.039}^{+0.047}$ \\
		$f_g$		& 0.021$_{-0.011}^{+0.019}$ & 0.040$_{-0.017}^{+0.025}$  & 0.077$_{-0.027}^{+0.035}$ & 0.042$_{-0.020}^{+0.029}$ & 0.060$_{-0.025}^{+0.034}$ \\
		$f_r$		& 0.941$_{-0.016}^{+0.007}$ & 0.939$_{-0.021}^{+0.010}$  & 0.921$_{-0.040}^{+0.023}$ & 0.935$_{-0.025}^{+0.013}$ & 0.884$_{-0.066}^{+0.047}$ \\
		                &                           &                            &                           &                           &                           \\
                                & \multicolumn{5}{c}{$M_{\star} < 10^{10.5}M_{\odot}$}                                                                                       \\
		                &                           &                            &                           &                           &                           \\
		$f_b$		& 0.089$_{-0.030}^{+0.046}$ & 0.336$_{-0.044}^{+0.046}$  & 0.393$_{-0.042}^{+0.045}$ & 0.460$_{-0.042}^{+0.044}$ & 0.553$_{-0.038}^{+0.038}$ \\
		$f_g$		& 0.065$_{-0.024}^{+0.029}$ & 0.102$_{-0.025}^{+0.029}$  & 0.128$_{-0.028}^{+0.030}$ & 0.120$_{-0.023}^{+0.026}$ & 0.133$_{-0.022}^{+0.024}$ \\
		$f_r$		& 0.882$_{-0.047}^{+0.040}$ & 0.647$_{-0.052}^{+0.051}$  & 0.534$_{-0.047}^{+0.046}$ & 0.473$_{-0.040}^{+0.041}$ & 0.363$_{-0.034}^{+0.035}$ \\
		\hline
		\hline 
		Substructures   &			    &                            &                           &                           &                           \\
                                & \multicolumn{5}{c}{$M_{\star} \geq 10^{10.5}M_{\odot}$}                                                                                    \\
		                &                           &                            &                           &                           &                           \\
		$f_b$		& 0.204$_{-0.069}^{+0.136}$ & 0.197$_{-0.050}^{+0.100}$  & 0.434$_{-0.114}^{+0.133}$ & 0.181$_{-0.039}^{+0.079}$ & 0.266$_{-0.095}^{+0.153}$ \\
		$f_g$		& 0.095$_{-0.069}^{+0.136}$ & 0.068$_{-0.050}^{+0.100}$  & 0.046$_{-0.033}^{+0.070}$ & 0.134$_{-0.092}^{+0.155}$ & 0.057$_{-0.042}^{+0.086}$ \\
		$f_r$		& 0.640$_{-0.346}^{+0.171}$ & 0.759$_{-0.131}^{+0.064}$  & 0.647$_{-0.179}^{+0.139}$ & 0.754$_{-0.129}^{+0.068}$ & 0.732$_{-0.182}^{+0.088}$ \\
		                &                           &                            &                           &                           &                           \\
                                & \multicolumn{5}{c}{$M_{\star} < 10^{10.5}M_{\odot}$}                                                                                    \\
		                &                           &                            &                           &                           &                           \\
		$f_b$		& 0.712$_{-0.164}^{+0.165}$ & 0.632$_{-0.065}^{+0.064}$  & 0.598$_{-0.048}^{+0.048}$ & 0.620$_{-0.042}^{+0.040}$ & 0.611$_{-0.039}^{+0.040}$ \\
		$f_g$		& 0.056$_{-0.038}^{+0.063}$ & 0.119$_{-0.034}^{+0.041}$  & 0.096$_{-0.024}^{+0.028}$ & 0.094$_{-0.021}^{+0.025}$ & 0.068$_{-0.017}^{+0.021}$ \\
		$f_r$		& 0.389$_{-0.103}^{+0.108}$ & 0.294$_{-0.054}^{+0.058}$  & 0.326$_{-0.044}^{+0.045}$ & 0.301$_{-0.038}^{+0.039}$ & 0.333$_{-0.037}^{+0.038}$ \\

		\hline
		\hline
		\end{tabular}  
}
\end{threeparttable}
\end{table*}

\begin{table*}
\centering
\begin{threeparttable}
	\caption{Spectrophotometric colour fraction as a function of $R_{c}$}
  	\label{tablea11}
	{\small
 		\begin{tabular}{lcccccc}
		\hline
		\hline
 			        & \multicolumn{5}{c}{$R_{c}$}         		                                                                                     \\
				& 19.5                      & 20.5                        & 21.5                       & 22.5                       & 23.5                       \\
		\hline
		Cluster         &                           &                            &                           &                           &                           \\
		$f_b$		& 0.298$_{-0.081}^{+0.087}$ & 0.255$_{-0.019}^{+0.021}$  & 0.370$_{-0.016}^{+0.016}$ & 0.468$_{-0.017}^{+0.017}$ & 0.656$_{-0.017}^{+0.017}$ \\
		$f_g$		& 0.118$_{-0.037}^{+0.044}$ & 0.084$_{-0.013}^{+0.015}$  & 0.119$_{-0.011}^{+0.012}$ & 0.139$_{-0.012}^{+0.012}$ & 0.127$_{-0.010}^{+0.011}$ \\
		$f_r$		& 0.605$_{-0.071}^{+0.069}$ & 0.625$_{-0.023}^{+0.021}$  & 0.497$_{-0.019}^{+0.018}$ & 0.386$_{-0.018}^{+0.017}$ & 0.206$_{-0.014}^{+0.014}$ \\
		\hline
		\hline 
		Substructures   &			    &                            &                           &                           &                           \\
		$f_b$		& 0.205$_{-0.061}^{+0.122}$ & 0.212$_{-0.044}^{+0.064}$  & 0.353$_{-0.052}^{+0.056}$ & 0.289$_{-0.055}^{+0.058}$ & 0.394$_{-0.055}^{+0.055}$ \\
		$f_g$		& 0.084$_{-0.061}^{+0.122}$ & 0.025$_{-0.018}^{+0.038}$  & 0.085$_{-0.028}^{+0.035}$ & 0.147$_{-0.033}^{+0.037}$ & 0.144$_{-0.030}^{+0.034}$ \\
		$f_r$		& 0.714$_{-0.224}^{+0.105}$ & 0.792$_{-0.065}^{+0.034}$  & 0.561$_{-0.059}^{+0.057}$ & 0.480$_{-0.051}^{+0.049}$ & 0.371$_{-0.045}^{+0.045}$ \\
		\hline
		\hline
		\end{tabular}  
}
\end{threeparttable}
\end{table*}

\begin{table*}
\centering
\begin{threeparttable}
	\caption{Spectrophotometric colour fraction as a function of $\log(M_{\star}/M_{\odot})$}
  	\label{tablea12}
	{\small
 		\begin{tabular}{lcccc}
		\hline
		\hline
		                & \multicolumn{4}{c}{$\log(M_{\star}/M_{\odot})$}		                                                \\
		\hline
				& 8.5	                    & 9.5                 	& 10.5	                    & 11.5	        	\\
		\hline
		Cluster 	&                           &                           &                           &		                \\
		$f_b$		& 0.860$_{-0.011}^{+0.010}$ & 0.626$_{-0.012}^{+0.011}$ & 0.301$_{-0.011}^{+0.010}$ & 0.127$_{-0.011}^{+0.020}$	\\
		$f_g$		& 0.069$_{-0.007}^{+0.007}$ & 0.137$_{-0.008}^{+0.008}$ & 0.112$_{-0.008}^{+0.009}$ & 0.100$_{-0.022}^{+0.025}$	\\
		$f_r$		& 0.071$_{-0.007}^{+0.008}$ & 0.227$_{-0.010}^{+0.010}$ & 0.565$_{-0.013}^{+0.013}$ & 0.652$_{-0.034}^{+0.031}$	\\
		\hline
		\hline		
		Substructures	&                           &                           &                           &		                \\
		$f_b$		& 0.844$_{-0.029}^{+0.027}$ & 0.440$_{-0.032}^{+0.034}$ & 0.277$_{-0.030}^{+0.033}$ & 0.305$_{-0.081}^{+0.102}$	\\
		$f_g$		& 0.051$_{-0.013}^{+0.016}$ & 0.132$_{-0.020}^{+0.022}$ & 0.072$_{-0.021}^{+0.025}$ & 0.039$_{-0.028}^{+0.059}$	\\
		$f_r$		& 0.104$_{-0.020}^{+0.023}$ & 0.394$_{-0.031}^{+0.031}$ & 0.636$_{-0.042}^{+0.040}$ & 0.613$_{-0.138}^{+0.107}$	\\
		\hline
		\hline
		\end{tabular}  
}
\end{threeparttable}
\end{table*}

\begin{table*}
\centering
\begin{threeparttable}
	\caption{Mean environmental quenching efficiency as a function of distance from the cluster centre for the spectroscopic sample.}
  	\label{tablea13}
	{\small
 		\begin{tabular}{lccccc}
		\hline
		\hline

		$r_{cl}/r_{200_{cl}}$		& 0.5                       & 1.5                        & 2.5                       & 3.5                        \\
 			& \multicolumn{4}{c}{$\epsilon_q$}         		       \\
		\hline
		Cluster Galaxies			& 0.709$_{-0.008}^{+0.008}$ & 0.151$_{-0.011}^{+0.011}$  & 0.012$_{-0.018}^{+0.019}$ & 0.382$_{-0.148}^{+0.150}$ \\
		Massive Cluster Galaxies		& 0.768$_{-0.024}^{+0.024}$ & 0.021$_{-0.076}^{+0.075}$  & 0.805$_{-0.093}^{+0.094}$ & 0.266$_{-0.194}^{+0.300}$ \\
		Less Massive Cluster Galaxies 		& 0.540$_{-0.010}^{+0.010}$ & 0.078$_{-0.010}^{+0.010}$  & 0.031$_{-0.015}^{+0.015}$ & 0.023$_{-0.061}^{+0.086}$ \\
		\hline
		\hline
		Substructure Galaxies			& 0.593$_{-0.029}^{+0.029}$ & 0.086$_{-0.020}^{+0.019}$  & 0.066$_{-0.048}^{+0.051}$ & -                         \\
		Massive Substructure Galaxies		& 0.600$_{-0.078}^{+0.078}$ & 0.056$_{-0.207}^{+0.206}$  & 0.171$_{-0.514}^{+0.572}$ & -                         \\
		Less Massive Substructure Galaxies	& 0.426$_{-0.035}^{+0.035}$ & 0.020$_{-0.008}^{+0.008}$  & 0.096$_{-0.021}^{+0.022}$ & -                         \\
		\hline
		\hline
		\end{tabular}  
}
\end{threeparttable}
\end{table*}

\begin{table*}
\centering
\begin{threeparttable}
	\caption{Mean environmental quenching efficiency as a function of distance from the overdensity centre for the spectroscopic sample.}
  	\label{tablea14}
	{\small
 		\begin{tabular}{lcccccc}
		\hline
		\hline
		$r/r_{200}$     & 0.1                       & 0.3                        & 0.5                       & 0.7                       & 0.9                       \\
 			        & \multicolumn{5}{c}{$\epsilon_q$}         		                                                                                     \\
		\hline
		Cluster Galaxies			& 0.875$_{-0.014}^{+0.014}$ & 0.825$_{-0.015}^{+0.015}$  & 0.728$_{-0.025}^{+0.025}$ & 0.593$_{-0.028}^{+0.028}$ & 0.474$_{-0.028}^{+0.028}$ \\
		Massive Cluster Galaxies		& 0.903$_{-0.032}^{+0.032}$ & 0.836$_{-0.039}^{+0.039}$  & 0.685$_{-0.041}^{+0.040}$ & 0.800$_{-0.058}^{+0.058}$ & 0.420$_{-0.100}^{+0.100}$ \\
		Less Massive Cluster Galaxies		& 0.741$_{-0.021}^{+0.021}$ & 0.635$_{-0.024}^{+0.024}$  & 0.558$_{-0.034}^{+0.034}$ & 0.407$_{-0.033}^{+0.033}$ & 0.386$_{-0.021}^{+0.021}$ \\
		\hline
		\hline 
		Substructure Galaxies			& 0.361$_{-0.053}^{+0.053}$ & 0.418$_{-0.044}^{+0.044}$  & 0.520$_{-0.048}^{+0.048}$ & 0.445$_{-0.041}^{+0.041}$ & 0.006$_{-0.018}^{+0.018}$ \\
		Massive Substructure Galaxies		& 0.387$_{-0.131}^{+0.194}$ & 0.772$_{-0.183}^{+0.181}$  & 0.659$_{-0.236}^{+0.241}$ & 0.721$_{-0.210}^{+0.206}$ & 0.234$_{-0.207}^{+0.267}$ \\
		Less Massive Substructure Galaxies	& 0.239$_{-0.053}^{+0.054}$ & 0.194$_{-0.043}^{+0.043}$  & 0.298$_{-0.054}^{+0.054}$ & 0.222$_{-0.042}^{+0.042}$ & 0.106$_{-0.011}^{+0.011}$ \\
		\hline
		\hline
		\end{tabular}  
}
\end{threeparttable}
\end{table*}

\begin{table*}
\centering
\begin{threeparttable}
	\caption{Mean environmental quenching efficiency as a function of distance from the cluster centre for the spectrophotometric sample.}
  	\label{tablea15}
	{\small
 		\begin{tabular}{lccccc}
		\hline
		\hline

		$r_{cl}/r_{200_{cl}}$		& 0.5                       & 1.5                        & 2.5                       & 3.5                        \\
 			& \multicolumn{4}{c}{$\epsilon_q$}         		       \\
		\hline
		Cluster Galaxies			& 0.676$_{-0.013}^{+0.013}$ & 0.136$_{-0.011}^{+0.011}$  & 0.091$_{-0.010}^{+0.010}$ & 0.147$_{-0.017}^{+0.017}$ \\
		Massive Cluster Galaxies		& 0.844$_{-0.031}^{+0.031}$ & 0.102$_{-0.076}^{+0.075}$  & 0.180$_{-0.080}^{+0.080}$ & 0.450$_{-0.079}^{+0.101}$ \\
		Less Massive Cluster Galaxies 		& 0.548$_{-0.016}^{+0.016}$ & 0.122$_{-0.011}^{+0.011}$  & 0.097$_{-0.009}^{+0.009}$ & 0.176$_{-0.017}^{+0.017}$ \\
		\hline
		\hline
		Substructure Galaxies			& 0.570$_{-0.030}^{+0.030}$ & 0.188$_{-0.023}^{+0.023}$  & 0.071$_{-0.054}^{+0.069}$ & -                         \\
		Massive Substructure Galaxies		& 0.930$_{-0.062}^{+0.061}$ & 0.185$_{-0.175}^{+0.194}$  & 0.670$_{-0.742}^{+0.745}$ & -                         \\
		Less Massive Substructure Galaxies	& 0.535$_{-0.030}^{+0.030}$ & 0.204$_{-0.021}^{+0.022}$  & 0.006$_{-0.059}^{+0.081}$ & -                         \\
		\hline
		\hline
		\end{tabular}  
}
\end{threeparttable}
\end{table*}

\begin{table*}
\centering
\begin{threeparttable}
	\caption{Mean environmental quenching efficiency as a function of distance from the overdensity centre for the spectrophotometric sample.}
  	\label{tablea16}
	{\small
 		\begin{tabular}{lcccccc}
		\hline
		\hline
		$r/r_{200}$     & 0.1                       & 0.3                        & 0.5                       & 0.7                       & 0.9                       \\
 			        & \multicolumn{5}{c}{$\epsilon_q$}         		                                                                                     \\
		\hline
		Cluster Galaxies			& 0.999$_{-0.024}^{+0.023}$ & 0.849$_{-0.029}^{+0.029}$  & 0.698$_{-0.033}^{+0.032}$ & 0.628$_{-0.031}^{+0.031}$ & 0.044$_{-0.029}^{+0.029}$ \\
		Massive Cluster Galaxies		& 0.972$_{-0.025}^{+0.025}$ & 0.963$_{-0.034}^{+0.034}$  & 0.905$_{-0.069}^{+0.069}$ & 0.951$_{-0.042}^{+0.042}$ & 0.789$_{-0.123}^{+0.124}$ \\
		Less Massive Cluster Galaxies		& 0.999$_{-0.039}^{+0.039}$ & 0.733$_{-0.046}^{+0.046}$  & 0.584$_{-0.042}^{+0.042}$ & 0.504$_{-0.036}^{+0.037}$ & 0.359$_{-0.031}^{+0.031}$ \\
		\hline
		\hline 
		Substructure Galaxies			& 0.537$_{-0.093}^{+0.091}$ & 0.361$_{-0.053}^{+0.052}$  & 0.368$_{-0.041}^{+0.040}$ & 0.336$_{-0.036}^{+0.036}$ & 0.377$_{-0.035}^{+0.035}$ \\
		Massive Substructure Galaxies		& 0.298$_{-0.284}^{+0.282}$ & 0.740$_{-0.214}^{+0.214}$  & 0.323$_{-0.176}^{+0.231}$ & 0.718$_{-0.216}^{+0.216}$ & 0.638$_{-0.281}^{+0.300}$ \\
		Less Massive Substructure Galaxies	& 0.468$_{-0.094}^{+0.094}$ & 0.324$_{-0.050}^{+0.050}$  & 0.373$_{-0.040}^{+0.040}$ & 0.335$_{-0.035}^{+0.034}$ & 0.383$_{-0.033}^{+0.034}$ \\
		\hline
		\hline
		\end{tabular}  
}
\end{threeparttable}
\end{table*}

\clearpage
								
\end{document}